\documentclass[draft]{agujournal2019}
\usepackage{url} 
\usepackage[inline]{trackchanges} 
\usepackage{soul}
\usepackage{amsmath}
\draftfalse

\journalname{AGU Advances}

\begin{document}

%
%


\title{Radiative and dynamical controls on the land-ocean warming contrast in climate models}

%
%




\authors{Paolo Giani\affil{1}, Arlene M. Fiore\affil{1}, Raffaele Ferrari\affil{1}, Paul A. O'Gorman\affil{1}, Vincent T. Cooper\affil{1}, Noelle E. Selin\affil{1,2,3} }

\affiliation{1}{Department of Earth, Atmospheric, and Planetary Sciences, Massachusetts Institute of Technology, Cambridge, Massachusetts}
\affiliation{2}{Center for Sustainability, Science and Strategy, Massachusetts Institute of Technology, Cambridge, Massachusetts}
\affiliation{3}{Institute for Data, Systems, and Society, Massachusetts Institute of Technology, Cambridge, Massachusetts}




\correspondingauthor{Paolo Giani}{pgiani@mit.edu}


\begin{keypoints}
\item Energetic and dynamical explanations of the land-ocean warming contrast reflect similar underlying physics
\item The warming contrast emerges from a model-dependent radiative baseline and a robust dynamical restoring mechanism that favors land warming
\item We clarify how climate sensitivity relates to the strength of the land-ocean warming contrast
\end{keypoints}
%
%

%
%


\begin{abstract}
Surface air over land warms substantially more than over the ocean under greenhouse forcing, a phenomenon known as the land--ocean warming contrast. Current explanations for this contrast are commonly expressed either in terms of energetic constraints, from top-of-atmosphere and surface energy balance, or dynamical constraints, from large-scale atmospheric dynamics. We show that these perspectives are complementary when viewed through the lens of atmospheric moist static energy (MSE) transport, and that connecting them yields new insight into the controls of the warming contrast and the spread in climate models. We use this framework to construct an interpretable emulator that reproduces the land--ocean warming response across 22 models from the latest Coupled Model Intercomparison Project (CMIP6). We find that the strength of the land--ocean warming contrast emerges from the interplay between a model-dependent radiative baseline and a robust dynamical restoring mechanism that favors greater warming over land. This interplay leads to two broad model regimes that align with climate sensitivity. In low-climate-sensitivity models, more stabilizing radiative feedbacks over the ocean directly favor greater land warming. In high-climate-sensitivity models, radiative feedbacks alone would instead favor greater ocean warming, but a strong MSE-transport feedback (approximately 0.2 PW K$^{-1}$) more than compensates for this tendency. The intermodel spread in the land--ocean warming contrast is closely related to the ratio of radiative feedbacks over land and ocean, highlighting a broader connection between climate sensitivity and the land--ocean warming contrast.
\end{abstract}

\section*{Plain Language Summary}
A notable feature of current and projected climate change is that land areas tend to warm more than ocean areas, although the strength of this contrast varies across climate models. Two main theories have been proposed to explain this land--ocean warming contrast, based either on principles of large-scale atmospheric dynamics or on energy balance. In this work, we show that these two perspectives are essentially two sides of the same coin and reflect similar underlying physics (associated with differences in relative humidity over land and ocean). We use this connection to construct a simple model that reproduces the behavior of much more complex climate models while remaining easy to interpret. This allows us to isolate the effects of processes such as differences in land and ocean heat capacity, radiative forcing and feedbacks (for example, changes in cloud properties with warming), and atmospheric energy transport. We find that differences in radiative feedbacks across climate models, which are also important for explaining why models warm by different amounts under the same increase in CO$_2$, play a central role in determining why models simulate different strengths of the land--ocean warming contrast. We also quantify how radiative processes and atmospheric energy transport interact to produce greater warming over land.
%
%

%


%
%
%
%

\section{Introduction}

A striking feature of both present-day and projected climate change is the spatially inhomogeneous pattern of warming. At leading order, this pattern is dominated by the land-ocean warming contrast, whereby surface air temperatures over land increase more than over the ocean. The land-ocean warming contrast is a robust response to changes in greenhouse gas concentrations, and is supported by evidence from observations \cite{byrneTrendsContinentalTemperature2018, wallaceComparisonLandOcean2018}, models \cite{todaEnergyBudgetFramework2023}, and paleoclimate proxies from the last glacial maximum \cite{seltzerTerrestrialAmplificationPresent2023}. Several studies have shown that this contrast is not only a transient feature due to the larger heat capacity of the ocean, but also an equilibrium response, as demonstrated in slab ocean simulations \cite{suttonLandSeaWarming2007} and fixed-SST experiments \cite{dommengetOceansRoleContinental2009}. The strength of the warming contrast is typically quantified by the land-ocean amplification factor $\phi$, defined as the ratio between mean land and ocean surface air temperature anomalies, and varies across different climate models.

What sets the amplification factor $\phi$, and why does it differ across models? Several theories have been proposed to explain why $\phi > 1$, which can be broadly grouped into two categories: (1) a dynamical perspective, in which the atmosphere provides a constraint through moist static energy (MSE) exchange across land and ocean regions \cite{joshiMechanismsLandSea2008, byrneLandOceanWarming2013, byrneUnderstandingDecreasesLand2016, byrneTrendsContinentalTemperature2018}, and (2) an energetic perspective, in which top-of-atmosphere (TOA) and surface energy balance over land and ocean regions provide a diagnostic framework to interpret the warming contrast \cite{lambertControlLandoceanTemperature2007, geoffroyLandseaWarmingContrast2015, todaEnergyBudgetFramework2021, todaEnergyBudgetFramework2023}.  
The dynamical perspective is based on the idea that large-scale atmospheric circulation couples surface air temperatures over land and ocean, preventing them from changing independently. This is often referred to as the lapse-rate mechanism because the warming contrast arises from differences in the moist adiabatic lapse rates, which in turn are related to the contrast in surface relative humidity between land and ocean. The assumptions of the lapse-rate mechanism (weak horizontal temperature gradient in the free troposphere, and convective quasi-equilibrium, \citeA{byrneLandOceanWarming2013}) are mostly valid in the tropics. One limitation of the dynamical perspective is that most land areas are in the extratropics, where horizontal temperature gradients can be sustained through geostrophic balance. Moreover, while the dynamical perspective is effective at explaining why $\phi > 1$ in the tropics, we find it offers limited insight into intermodel differences.

The energetic perspective is presented in \citeA{suttonLandSeaWarming2007}, who argued that surface forcing is primarily balanced by latent heat flux over the ocean (through evaporation), but not over land, where moisture is more limited. This is sometimes referred to as the Bowen ratio mechanism. This surface-based framework was later extended to a TOA perspective \cite{lambertRelationshipLandOcean2011, todaEnergyBudgetFramework2021, todaEnergyBudgetFramework2023}, where the contrast was linked to horizontal energy transport in the atmosphere \cite{geoffroyLandseaWarmingContrast2015}, as well as differences in radiative forcing between land and ocean regions \cite{todaEnergyBudgetFramework2021}. \citeA{todaEnergyBudgetFramework2021} also showed that the intermodel spread in individual contributions (e.g., radiative feedbacks, effective radiative forcing) is much larger than the spread in $\phi$, implying substantial compensation among different processes. 

While offering a useful framework to interpret the land-ocean warming contrast and the intermodel spread, it is not obvious how the energy balance framework relates to the dynamical perspective.
In this work, we explicitly link the energetic and dynamical perspectives and show that the two views are complementary (Section \ref{sec:connection}). We use this connection to develop an energy-balance-based interpretable emulator \cite{tebaldiEmulatorsClimateModel2025a} that provides a minimal representation of the physics required to capture the land-ocean warming contrast in climate models (Section \ref{sec:emulator}). We analyze the emulator parameters across 22 climate models from the latest Coupled Model Intercomparison Project (CMIP6) to understand what drives the intermodel spread in $\phi$, and to connect the dynamical and energetic explanations found in previous literature. Using this framework, we show that radiative feedback differences set a baseline $\phi$ that individual models would favor in the absence of atmospheric heat transport, while large-scale atmospheric dynamics acts as a restoring mechanism that shifts the response toward greater warming over land and reduces the intermodel spread in $\phi$. This framework explains why all models tend to produce $\phi>1$ despite large differences in their radiative feedback configuration. It also helps explain the moderate correlation between equilibrium climate sensitivity (ECS) and $\phi$ noted in \citeA{todaEnergyBudgetFramework2021}.

\section{Connecting the dynamical and energetic perspectives on the land-ocean warming contrast}

\label{sec:connection}

In this Section, we briefly review the dynamical and energetic perspectives before drawing an explicit connection between the two, based on moist static energy exchange. As noted by \citeA{todaEnergyBudgetFramework2021}, the dynamical and energetic perspectives, as well as the different heat transport parameterizations, seem to offer different interpretations as to the factors driving $\phi > 1$. In the next section, however, we show that these approaches are closely connected within the MSE framework, and that combining them yields new insights into several aspects of the land-ocean warming contrast.

\subsection{The dynamical perspective}

The dynamical perspective seeks to explain the equilibrium land-ocean warming contrast using constraints from atmospheric dynamics. It relies on two main assumptions \cite{joshiMechanismsLandSea2008, byrneLandOceanWarming2013}: (1) on average, the vertical structure of the atmosphere over both land and ocean follows a moist adiabat, and (2) temperatures over land and ocean are equal sufficiently high in the atmosphere. These arguments hold approximately in the tropics \cite{byrneTrendsContinentalTemperature2018}. Under these two assumptions, \citeA{byrneLandOceanWarming2013} showed that, at equilibrium, surface MSE over land is equal to surface MSE over the ocean, and that this equality is maintained following a climate perturbation:
\begin{equation}
\Delta \text{MSE}_L( \Delta q_L, \Delta T_L) = \Delta \text{MSE}_O(\Delta q_O, \Delta T_O)
\label{eq:MSEconstraint}
\end{equation}
The subscripts $L$ and $O$ imply averaging over land and ocean areas, respectively, $\Delta \text{MSE} = c_p\Delta T + L_v\Delta q$, $q$ is surface air specific humidity, $T$ is surface air temperature, $L_v$ is the latent heat of vaporization (2.5 MJ kg$^{-1}$) and $c_p$ is the constant-pressure specific heat capacity of air (1005 J kg$^{-1}$ K$^{-1}$).

To use Equation \ref{eq:MSEconstraint} as a diagnostic for the land-ocean amplification factor $\phi$, an additional constraint is needed to specify how land and ocean moisture respond to a climate perturbation ($\Delta q_L$ and $\Delta q_O$). \citeA{byrneUnderstandingDecreasesLand2016} assumed that surface relative humidity over the ocean ($r_O$) is climate-invariant, such that $q_O$ increases with temperature according to the Clausius--Clapeyron relationship ($\Delta q_O = q_O\alpha \Delta T_O$, where $\alpha = \partial \log(q_{\mathrm{sat}})/\partial T_O$ is the fractional sensitivity of saturation specific humidity $q_{\mathrm{sat}}$ to temperature). They further assumed that $q_L$ is tied to $q_O$ through a climate-invariant factor $\gamma$ (reflecting the predominantly oceanic source of atmospheric moisture), such that $q_L = \gamma q_O$. Under these assumptions, relative humidity decreases over land and Equation \ref{eq:MSEconstraint} yields:
\begin{equation}
\Delta T_L = \Delta T_O\left( 1 + \frac{L_v}{c_p}\alpha(1 - \gamma)q_O\right).
\label{eq:phi-dynamical}
\end{equation}
For typical values of $T_O = 289$ K, $r_O = 80$\%, $\alpha = 7\%/K$  and $\gamma = 0.70$ \cite{byrneUnderstandingDecreasesLand2016,byrneTrendsContinentalTemperature2018}, and a perturbation of $\Delta T_O = 2$ K, Equation \ref{eq:phi-dynamical} predicts $\Delta T_L = 2.9$ K and $\phi = 1.46$, qualitatively consistent with models.

\subsection{The energetic perspective}

The energetic perspective explains the land-ocean warming contrast using local energy balance arguments. The central idea is the TOA forcing–feedback framework that governs global-mean climate sensitivity \cite{gregoryNewMethodDiagnosing2004}, applied locally \cite{armourTimeVaryingClimateSensitivity2013, gianiOriginLimitsInvariant2026} to atmospheric columns over land and ocean (accounting for heat transport changes).  Following a radiative perturbation, the energy content of each column evolves in response to changes in local radiative properties, which is typically decomposed into a forcing term $F$ and a feedback term $\lambda \Delta T$ (where $\lambda$ is a radiative feedback parameter), as well as to changes in heat transport between the columns:
\begin{equation}
\begin{aligned}
\Delta U_L &= \Delta N_L + \Delta H/a_L
          = F_L - \lambda_L \Delta T_L + \Delta H/a_L \\
\Delta U_O &= \Delta N_O - \Delta H/a_O
          = F_O - \lambda_O \Delta T_O - \Delta H/a_O
\end{aligned}
\label{eq:energybalance}
\end{equation}
Here $\Delta U$ is the net surface heat uptake, $\Delta N$ is the change in TOA radiative flux (positive downward), $\Delta H$ is the change in energy transport in J s$^{-1}$ between the columns (defined positive as transporting energy from ocean to land, i.e. a positive value implies a warming tendency for land and a cooling tendency for the ocean), and $a_i$ is the surface area of the column, such that all values are expressed in W m$^{-2}$, and  $i \in \{L, O\}$ indexes the two columns. We define $\lambda > 0$ as stabilizing. At equilibrium, $\Delta U_i = 0$ (i.e., the system does not accumulate energy), and Equation \ref{eq:energybalance} provides a diagnostic relationship for $\Delta T_L$ and $\Delta T_O$, given a closure parameterization for $\Delta H$ and specified values of $F_i$ and $\lambda_i$. 

\citeA{todaEnergyBudgetFramework2021} and \citeA{todaEnergyBudgetFramework2023} showed that a reasonable energy transport closure parameterization for a radiative perturbation (e.g., 4$\times$CO$_2$) is:
\begin{equation}
    \Delta H = F_H + \lambda_H \Delta T_O
    \label{eq:toda-param}
\end{equation}
i.e., a fast adjustment $F_H$ in the climatological ocean-to-land energy transport (which can be viewed as a ``forcing") and a slower response mediated by $\Delta T_O$ with linear coefficient $\lambda_H$ (which can be viewed as a ``feedback"). With this parameterization, \citeA{todaEnergyBudgetFramework2021} decomposed the equilibrium amplification factor $\phi$ into contributions from radiative forcing, radiative feedbacks, and changes in heat transport across CMIP6 models. They found that the intermodel spread in each individual contribution is large compared to the spread in $\phi$, indicating substantial compensation among processes. \citeA{geoffroyLandseaWarmingContrast2015} showed that a purely diffusive parameterization in temperature (e.g., $\Delta H = -D(\Delta T_L - \Delta T_O)$) is not suitable and proposed the following parameterization:
\begin{equation}
    \Delta H = -(k_L\Delta T_L - k_O\Delta T_O)
    \label{eq:geoff}
\end{equation}
where $k_L, k_O > 0$ are empirical coefficients describing the sensitivity of horizontal energy transport to warming over land and ocean, respectively. Under our sign convention, land warming drives transport from land to ocean, whereas ocean warming drives transport from ocean to land.

\subsection{Connecting the two perspectives} 

To connect the energetic and dynamical perspectives, we begin by considering two limiting cases and then show that these different perspectives and parameterizations can be reconciled within a framework of downgradient surface MSE exchange between land and ocean that lies between these two limits.

The dynamical perspective corresponds to the limit in which the atmosphere is perfectly efficient at exchanging MSE between land and ocean (Equation \ref{eq:MSEconstraint}). In this limit, $\Delta H$ is diagnosed from the constraint $\Delta \mathrm{MSE}_L = \Delta \mathrm{MSE}_O$ while simultaneously satisfying local energy balance. Combining the energetic relationship in Equation \ref{eq:energybalance}:
\begin{equation}
\Delta T_L = (F_L + \Delta H/a_L)/\lambda_L
\label{eq:energetic-relationship}
\end{equation}
with the MSE constraint from \citeA{byrneUnderstandingDecreasesLand2016} (Equation \ref{eq:phi-dynamical}), yields the following expression for $\Delta H$:
\begin{equation}
    \Delta H = -a_L(F_L - \lambda_L \phi \Delta T_O)
    \label{eq:inf-diffusivity}
\end{equation}
where $\phi = 1 + L_v\alpha (1 - \gamma)q_O/c_p$. This formulation reveals that the $\Delta H$ required to maintain $\Delta$MSE equality is linear in $\Delta T_O$, and the amplification ratio $\phi$ is independent of radiative feedbacks and instead controlled by $r_O$, $\gamma$ and $T_O$ (via the dependence of $q_{\text{sat}}$ on $T_O$). Radiative feedbacks determine the absolute temperature response to a given forcing, as well as the magnitude and sign of the energy transport $\Delta H$ required to maintain $\Delta$MSE equality. In this limit, there cannot be any land warming without ocean warming, and vice versa, because any MSE change is immediately communicated to the other region. This limit can also be interpreted as the infinite-diffusivity limit of downgradient MSE transport, $\Delta H = -D(\Delta \text{MSE}_L-\Delta \text{MSE}_O)$, where $D \rightarrow \infty$. As $D \rightarrow \infty$, the land-ocean MSE gradient approaches zero to maintain a finite $\Delta H$.  While \citeA{geoffroyLandseaWarmingContrast2015} briefly discuss the case of vanishing changes in horizontal energy transport ($\Delta H = 0$), the diffusive interpretation considered here allows a finite $\Delta H$, because $D \to \infty$ and $(\Delta \mathrm{MSE}_L-\Delta \mathrm{MSE}_O) \to 0$ can occur such that their product remains finite, as given by Equation \ref{eq:inf-diffusivity}. In other words, enforcing $\Delta$MSE equality does not imply $\Delta H = 0$.

The opposite limit corresponds to  uncoupled atmospheric columns, in which there is no energy exchange between land and ocean ($\Delta H = 0$). In this case, the land-ocean temperature response is determined only by local radiative forcing and feedbacks, yielding $\Delta T_L = F_L/\lambda_L$ and $\Delta T_O = F_O/\lambda_O$. The amplification ratio $\phi$ is therefore independent of $\gamma$, $T_O$ and $r_O$. Although this limit is not necessarily realistic, it serves as a useful baseline representing the opposite extreme of the dynamical perspective. 
A more realistic representation lies between these two limits, where the energetic and dynamical perspectives are linked by assuming that land and ocean regions exchange energy down the gradient of surface MSE, without requiring surface MSE changes to be identical over land and ocean.

To illustrate the intermediate case, we begin by showing that both \citeA{geoffroyLandseaWarmingContrast2015} and \citeA{todaEnergyBudgetFramework2021} parameterizations for heat transport are equivalent to downgradient exchange of surface MSE, i.e. $\Delta H = -D(\Delta \text{MSE}_L - \Delta \text{MSE}_O) = -D ( c_p\Delta T_L + L_v\Delta q_L - c_p\Delta T_O - L_v\Delta q_O )$, where $D$ is an effective atmospheric mass flux (units of kg s$^{-1}$) between land and ocean regions. Taking a similar approach to \citeA{geoffroyLandseaWarmingContrast2015}, but adding the moisture constraint from \citeA{byrneUnderstandingDecreasesLand2016}, we can write $\Delta q_O = q_O \alpha \Delta T_O$,  $\Delta q_L = \gamma\Delta q_O$ and:
\begin{equation} \Delta H = -D c_p\Delta T_L + D\left(c_p + (1-\gamma)L_vq_O\alpha\right)\Delta T_O \label{eq:MSEdiffusion} 
\end{equation} 
This is equivalent to the \citeA{geoffroyLandseaWarmingContrast2015} parameterization, where $k_L = Dc_p$ and $k_O = D\left(c_p + (1-\gamma)L_v q_O\alpha\right)$. This argument provides a simple explanation of what $k_L$ and $k_O$ physically represent, and it also highlights that $k_O > k_L$ as long as $\gamma < 1$, which explains the asymmetric heat transport argument described by \citeA{geoffroyLandseaWarmingContrast2015}, i.e. the fact that warming the ocean surface by 1 K transports more energy to land than an equivalent warming of 1 K over land. This also explains the finding of \citeA{geoffroyLandseaWarmingContrast2015} that, across CMIP5 models, the ratio between $k_L/k_O$ is much more constrained than their individual values, as the ratio does not depend on $D$, but only on $\alpha$, $r_O$, $T_O$ and $\gamma$. Physically, $k_O > k_L$ reflects that atmospheric energy transport is not downgradient in temperature, but instead approximately follows the gradient in MSE which is approximately conserved under moist-adiabatic motions (i.e., $k_L = k_O$ would imply downgradient temperature transport, e.g., in a dry atmosphere).

For the equivalence with the \citeA{todaEnergyBudgetFramework2021} formulation for a radiative perturbation (Equation \ref{eq:toda-param}), we note that, from the energetic perspective, temperature changes over land are governed by the sum of a radiative and a dynamical contribution (from Equation \ref{eq:energybalance}, assuming $\Delta U_L = 0$ because of small heat capacity):
\begin{equation}
    \Delta T_L =  \frac{F_L}{\lambda_L} + \frac{\Delta H/a_L}{\lambda_L}
\end{equation}

Using the $\Delta H$ formulation in Equation \ref{eq:MSEdiffusion} (surface MSE exchange) allows us to draw an equivalence between \citeA{geoffroyLandseaWarmingContrast2015} (Equation \ref{eq:geoff}) and \citeA{todaEnergyBudgetFramework2021}  (Equation \ref{eq:toda-param}) energy transport parameterizations: 

\begin{equation} F_H = -\frac{k_L F_L}{\left( \lambda_L + k_L/a_L \right)} 
\label{eq:fh} \end{equation} 
\begin{equation} \lambda_H = -\frac{k_L k_O/a_L}{\left(\lambda_L + k_L/a_L\right)} + k_O \label{eq:lh} \end{equation}

This framework provides a physical interpretation of the empirical parameters introduced by \citeA{todaEnergyBudgetFramework2021} in terms of MSE transport. The quantity $F_H$ (PW) represents the energy flux from land to ocean when $\Delta T_O = 0$ ($F_H <$ 0). Equation \ref{eq:fh} shows that $-F_H/a_L$ (in W m$^{-2}$) is a fraction $k_L/(a_L \lambda_L + k_L)$ of $F_L$, which is strictly less than 1 (approximately 80\% for typical values of $k_L$ and $\lambda_L$, as shown in the next Section).
In other words, a radiative perturbation over land induces a rapid adjustment in energy transport due to the fast warming of land (which increases MSE$_L$). This fast response accounts for roughly 80\% of the imposed land radiative forcing, with the exact value depending on MSE diffusivity (through $k_L=Dc_p)$ and land radiative feedbacks.
The parameter $\lambda_H$ represents the slower response and reflects the net effect of two competing processes: (i) a warmer ocean transports more energy to land (second term, $+k_O$), and (ii) a warmer ocean is associated with warmer land, which enhances energy transport from land to ocean (first term on the right-hand side of Equation \ref{eq:lh}). Due to the asymmetry between $k_O$ and $k_L$ ($k_O > k_L$, arising from $\gamma < 1$), the second term dominates, making $\lambda_H$ positive (ocean-to-land transport). This implies that each unit of $\Delta T_O$ leads to a net increase in energy transport towards land regions.

To summarize, this downgradient MSE exchange framework is more general than the more restrictive constraint of equal MSE, which is approximately valid only in the tropics. First, this framework allows for rapid land warming without concurrent ocean warming, which is not possible under the equal-MSE framework (and therefore enables a more realistic representation of transient scenarios such as abrupt-4xCO$_2$). Second, the equilibrium amplification ratio depends on radiative feedbacks in addition to moisture-related parameters, providing greater flexibility for interpreting intermodel differences in $\phi_{eq}$. We illustrate these advantages in the next Section by fitting a dynamical-system emulator to 22 CMIP6 models using the downgradient MSE parameterization. The goal is to identify the mechanisms through which different models produce the land-ocean warming contrast and to explain the spread in $\phi_{eq}$ together with other key features of the CMIP6 response.

\section{A prognostic dynamical system for the land-ocean warming contrast}
\label{sec:emulator}

\subsection{The dynamical system emulator}

We begin with a model similar to that of \citeA{geoffroyLandseaWarmingContrast2015}, but use Equation \ref{eq:toda-param} as the heat transport parameterization:

\begin{subequations}
\label{eq:emulator}
\begin{align}
C_O\frac{d\Delta T_O}{dt} &=
F_O - \lambda_O \Delta T_O
- \frac{F_H + \lambda_H \Delta T_O}{a_O}
- \epsilon\psi(\Delta T_O - \Delta T_D),
\label{eq:emulator-ocean}\\
C_D\frac{d\Delta T_D}{dt} &=
\psi(\Delta T_O - \Delta T_D),
\label{eq:emulator-deep}\\
\Delta T_L &=
\frac{F_L}{\lambda_L}
+ \frac{F_H + \lambda_H \Delta T_O}{a_L\lambda_L}.
\label{eq:emulator-land}
\end{align}
\end{subequations}
In Equation \ref{eq:emulator}, $C_O$ and $C_D$ are the mixed-layer and deep ocean heat capacities, respectively. Similarly to the traditional two-layer globally-averaged models for transient climate change \cite{heldProbingFastSlow2010, geoffroyTransientClimateResponse2013}, we model the deep ocean heat uptake as $K=\psi(\Delta T_O - \Delta T_D)$ (where $\psi$ is the heat exchange strength between surface and deep ocean, units W m$^{-2}$ K$^{-1}$), the TOA radiative flux over land as $\Delta N_L = F_L - \lambda_L\Delta T_L$, the radiative flux over the ocean as $\Delta N_O = F_O - \lambda_O\Delta T_O - (\epsilon - 1)K$ (the efficacy $\epsilon$ describes the pattern effect, i.e. the dependency of the ocean radiative feedback on the SST warming pattern, as in \cite{heldProbingFastSlow2010,wintonImportanceOceanHeat2010,geoffroyTransientClimateResponse2013a}). Note that the right-hand side of Equation \ref{eq:emulator-ocean} is the sum of $\Delta N_O$, $-\Delta H/a_O$, and the vertical ocean heat transport term $-K$. Although $\epsilon$ appears explicitly in the vertical transport term, it originates from the top-of-atmosphere radiative imbalance $\Delta N$ through the dependency of the effective radiative feedback on the deep ocean heat uptake $K$. The temperature response $\Delta T_L$ is diagnostic because land has a much smaller heat capacity, and we assume it immediately feels changes in radiative fluxes and horizontal energy transport. Equation \ref{eq:emulator} also allows us to directly determine changes in horizontal heat transport $\Delta H$ (according to Equation \ref{eq:toda-param}) and changes in TOA radiative fluxes.

\begin{figure}
    \centering
    \includegraphics[width=1.0\linewidth]{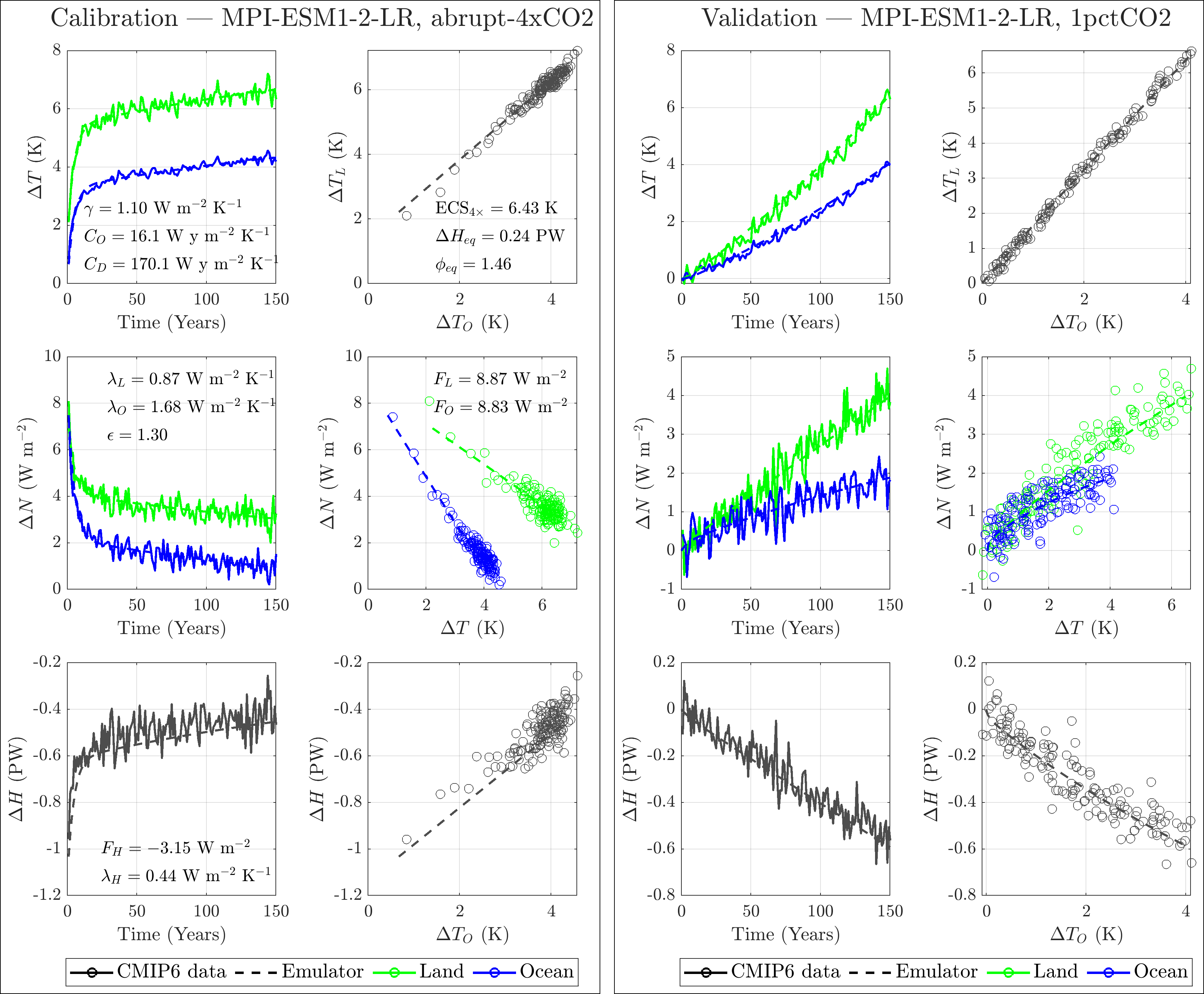}
    \caption{Comparison between the emulator described in Equation \ref{eq:emulator} and MPI-ESM1-2-LR data. The left half of the figure shows the abrupt-4xCO2 experiment, which is used to calibrate the 10 parameters that govern Equation \ref{eq:emulator}, and the right half shows 1pctCO2, which is used as out-of-sample validation. The top panels show the temperature responses $\Delta T$, the middle panels show changes in TOA radiative fluxes $\Delta N$, and the bottom panels show changes in heat transport $\Delta H$.}
    \label{fig:validation}
\end{figure}

Figure \ref{fig:validation} shows the solution of Equation \ref{eq:emulator} alongside model output from MPI-ESM1-2-LR, chosen as a representative model because its ECS is close to the CMIP6 multi-model average (approximately 3 K). The 10 parameters in Equation \ref{eq:emulator} ($C_O$, $C_D$, $F_{O,4\times}$, $F_{L,4\times}$, $\lambda_L$, $\lambda_O$, $\psi$, $\epsilon$, $F_{H,4\times}$, $\lambda_H$) are calibrated on the abrupt-4xCO2 experiment, and are kept constant for the 1pctCO2 validation in Figure \ref{fig:validation}. Calibration is achieved with simple linear regressions of the TOA radiative imbalance $\Delta N$ and net changes in heat transport $\Delta H$ against surface temperatures, along with an iterative procedure for the heat capacities, $\psi$ and $\epsilon$ (see Appendix B for more details on calibration/validation). Equation \ref{eq:emulator} not only reproduces the MPI-ESM1-2-LR surface air temperature response to increased CO$_2$, but also the response of dynamical and radiative heat fluxes, despite not capturing (by construction) internal variability. The fact that radiative fluxes, dynamical fluxes, and surface air temperatures are captured concurrently (i.e., they are mutually dependent and are not modeled separately) suggests that Equation \ref{eq:emulator} is able to capture key aspects of the dynamics represented in the MPI model. Perhaps more remarkably, Equation \ref{eq:emulator} also reproduces the behavior of the other 20+ CMIP6 models considered here reasonably well (Supplementary Material and Appendix A for more detail on the CMIP6 data), albeit with different calibrated parameters. This suggests that the physical mechanisms represented by the dynamical emulator are broadly robust across models, with much of the intermodel diversity captured through differences in the parameter values. This allows us to summarize the complexity of CMIP6 models with a common set of 10 parameters, enabling a much easier intercomparison that provides insight into the intermodel spread. 

Table \ref{tab:params} summarizes the parameter values for the 22 CMIP6 models that we consider. ECS$_{4\times}$ and $\phi_{eq}$ are diagnosed from the other parameters by solving Equation \ref{eq:emulator} at equilibrium.
\begin{table}
\caption{Radiative and dynamical parameter estimates for the 22 CMIP6 models that we consider in this work. Note that ECS$_{4\times}$ and $\phi_{eq}$ are derived from the other 8 parameters. The forcing terms ($F_{O,4\times}$, $F_{L,4\times}$) are in W m$^{-2}$, $F_{H,4\times}$ is in PW, the feedbacks $\lambda_L$, $\lambda_O$ and the deep ocean heat exchange strength $\psi$ are in W m$^{-2}$ K$^{-1}$; the feedback $\lambda_H$ is in PW K$^{-1}$; the efficacy of heat uptake $\epsilon$ is dimensionless, ECS$_{4\times}$ is in K and $\phi_{eq}$ is dimensionless.}
\renewcommand{\arraystretch}{1.4} 
\begin{tabular}{lllllllllll}
\hline
& $F_{O,4\times}$ & $F_{L,4\times}$ & $F_{H,4\times}$ & $\lambda_O$ & $\lambda_L$ & $\lambda_H$ & $\psi$ & $\epsilon$ & 
ECS$_{4\times}$ & $\phi_{eq}$\\ 
\hline
Ensemble Average & 7.26 & 8.09 & -0.99 & 0.87 & 0.96 & 0.17 & 1.00 & 1.36 & 8.31 & 1.40 \\ 
\hline
ACCESS-CM2 & 7.07 & 7.60 & -0.89 & 0.53 & 0.89 & 0.15 & 0.86 & 1.38 & 10.97 & 1.33 \\ 
AWI-CM-1-1-MR & 7.87 & 8.99 & -1.13 & 1.37 & 1.04 & 0.19 & 0.86 & 1.32 & 6.57 & 1.44 \\ 
BCC-CSM2-MR & 6.09 & 7.55 & -0.94 & 1.15 & 0.90 & 0.17 & 1.21 & 1.22 & 6.21 & 1.55 \\ 
CAS-ESM2-0 & 6.99 & 8.03 & -0.90 & 0.84 & 1.23 & 0.20 & 0.86 & 1.33 & 7.45 & 1.31 \\ 
CESM2 & 7.55 & 9.05 & -1.09 & 0.41 & 1.01 & 0.17 & 0.95 & 1.63 & 13.02 & 1.26 \\ 
CMCC-CM2-SR5 & 7.85 & 8.68 & -1.07 & 1.11 & 0.97 & 0.18 & 0.94 & 1.37 & 7.67 & 1.47 \\ 
CNRM-CM6-1 & 7.08 & 7.71 & -0.91 & 0.67 & 0.80 & 0.13 & 0.94 & 1.06 & 10.11 & 1.32 \\ 
CanESM5 & 7.49 & 7.74 & -0.96 & 0.57 & 0.77 & 0.14 & 0.85 & 1.11 & 11.75 & 1.40 \\ 
EC-Earth3 & 7.47 & 6.99 & -0.85 & 0.88 & 0.80 & 0.15 & 0.84 & 1.21 & 8.62 & 1.46 \\ 
FGOALS-f3-L & 9.31 & 8.17 & -1.00 & 1.51 & 1.15 & 0.21 & 1.01 & 1.46 & 6.52 & 1.46 \\ 
GFDL-CM4 & 8.49 & 6.92 & -0.89 & 0.91 & 0.63 & 0.12 & 1.00 & 1.88 & 9.95 & 1.39 \\ 
GISS-E2-1-G & 7.92 & 8.93 & -1.06 & 1.57 & 1.21 & 0.20 & 1.20 & 1.16 & 5.71 & 1.38 \\ 
HadGEM3-GC31-LL & 6.66 & 8.08 & -0.98 & 0.47 & 0.82 & 0.15 & 0.84 & 1.19 & 11.75 & 1.41 \\ 
INM-CM4-8 & 5.46 & 6.48 & -0.81 & 1.75 & 1.20 & 0.20 & 0.94 & 1.20 & 3.71 & 1.40 \\ 
IPSL-CM6A-LR & 7.18 & 7.35 & -0.85 & 0.64 & 0.85 & 0.15 & 0.62 & 1.33 & 10.06 & 1.42 \\ 
KACE-1-0-G & 7.10 & 6.87 & -0.89 & 0.52 & 0.81 & 0.16 & 1.49 & 1.38 & 11.23 & 1.44 \\ 
MIROC6 & 7.87 & 7.10 & -0.88 & 1.71 & 0.83 & 0.16 & 1.11 & 1.31 & 5.62 & 1.60 \\ 
MPI-ESM1-2-LR & 8.83 & 8.87 & -1.14 & 1.68 & 0.87 & 0.16 & 1.10 & 1.30 & 6.43 & 1.46 \\ 
MRI-ESM2-0 & 5.97 & 7.38 & -0.91 & 0.83 & 1.13 & 0.18 & 1.29 & 1.10 & 6.83 & 1.23 \\ 
NorESM2-LM & 10.00 & 8.21 & -1.05 & 1.57 & 0.99 & 0.20 & 1.26 & 2.42 & 7.07 & 1.55 \\ 
SAM0-UNICON & 7.85 & 10.75 & -1.25 & 0.89 & 1.35 & 0.21 & 1.01 & 1.44 & 8.29 & 1.25 \\ 
UKESM1-0-LL & 7.58 & 7.67 & -0.88 & 0.57 & 0.83 & 0.14 & 0.83 & 1.19 & 11.49 & 1.38 \\ 
\hline 
\end{tabular}
\label{tab:params}
\end{table}
As expected, there is substantial intermodel spread in the ocean radiative feedback parameter, $\lambda_O$. This parameter largely controls the spread in climate sensitivity and is closely tied to cloud feedbacks \cite{sherwoodAssessmentEarthClimate2020, bonyCloudsCirculationClimate2015, brethertonInsightsLowlatitudeCloud2015}. Despite the large variance in $\lambda_O$, the intermodel spread in the land-ocean amplification factor $\phi$ is relatively small, for reasons discussed in the next subsection. 
The relative strength of radiative feedbacks over land and ocean, $\lambda_L$ and $\lambda_O$, is not robust and varies substantially across models (i.e., roughly half of the models exhibit stronger radiative damping over the ocean $\lambda_O > \lambda_L$, while the other half exhibit stronger damping over land $\lambda_L > \lambda_O$). As a result, the multimodel mean shows little radiative feedback contrast, with $\lambda_L \approx \lambda_O$. 
Despite this lack of a consistent relationship between $\lambda_L$ and $\lambda_O$, all models produce $\phi_{eq} > 1$. This further confirms that a radiative feedback contrast favoring stronger land warming is not required to produce the land-ocean warming contrast, consistent with findings from \citeA{geoffroyLandseaWarmingContrast2015} based on CMIP5 models. The lack of model consensus in the relative strength of $\lambda_O$ and $\lambda_L$ is mostly due to the model spread in $\lambda_O$ (range 0.4---1.8 W m$^{-2}$ K$^{-1}$, standard deviation 0.44  W m$^{-2}$ K$^{-1}$ ), whereas $\lambda_L$ has a tighter range (0.6---1.3 W m$^{-2}$ K$^{-1}$, standard deviation 0.18 W m$^{-2}$ K$^{-1}$).

In contrast, the dynamical parameters $F_{H,{4\times}}$ (average -1.0 PW, range -1.2 to -0.8 PW) and $\lambda_H$ (average 0.17 PW K$^{-1}$, range 0.12 to 0.21 PW K$^{-1}$) are relatively more consistent across models. Their physical meanings can be directly connected to both the dynamical and energetic perspectives. The quantity $F_{H,4\times}$ represents the total energy transported from land to ocean following a 4$\times$CO$_2$ radiative perturbation (a positive $\Delta H$ corresponds to energy transport from ocean to land). From the perspective of MSE transport, $F_{H,4\times}$ arises from the rapid warming of land, which increases MSE over land before the ocean can adjust. This imbalance drives MSE transport from land to ocean. While there is some intermodel spread in $F_{H,4\times}$, it is strongly correlated with $F_L$ ($r = 0.9$), such that the ratio $F_H/F_L$ is more tightly constrained than either quantity individually. In particular, $-F_H/a_L$ is consistently about 80\% of $F_L$ (average 81.9\%, range 76 to 87\%). This suggests that the dynamical atmospheric response is robust across models, transporting roughly 80\% of the land radiative forcing to the ocean. The larger source of intermodel spread lies instead in the land radiative forcing, which largely determines the spread in $F_H$.

A similar result to the ``forcing" $F_{H,4\times} $ holds for the ``feedback"  $\lambda_H$, which quantifies the additional energy transport from ocean to land per unit increase in ocean temperature (approximately 0.17 PW K$^{-1}$). This transport can also be understood in terms of MSE exchange: a unit temperature increase in ocean temperature raises MSE over the ocean more than an equivalent increase over land, due to the larger moisture increase over the ocean at fixed relative humidity (since $r_O > r_L$). This effect occurs even if land relative humidity does not change, and is amplified if $r_L$ decreases as a function of warming. Much of the intermodel spread in $\lambda_H$ is explained by the spread in $\lambda_L$, as the two are strongly correlated ($r = 0.94$), reflecting the influence of $\lambda_L$ on $\Delta T_L$ (via the first term in Equation \ref{eq:lh}). Physically, a more stabilizing $\lambda_L$ (larger $\lambda_L$) implies a lower temperature increase over land for the same amount of energy, which reduces the MSE increase over land and increases the ocean-to-land energy transport (because of lower transport from land to ocean). Overall, these results indicate that the dynamical component of the models is relatively robust (models consistently act to transport MSE downgradient) while the primary source of uncertainty lies in the radiative parameters and their influence on the dynamical response.

The deep ocean heat exchange strength $\psi$ and the heat uptake efficacy $\epsilon$ are approximately 1 W m$^{-2}$ K$^{-1}$ and 1.3 (standard deviations 0.2 W m$^{-2}$ K$^{-1}$ and 0.3, respectively), consistent with values reported by \citeA{geoffroyTransientClimateResponse2013a}. Notably, $\epsilon > 1$ implies that the pattern effect acts to make the effective climate feedback less stabilizing in a warmer world, consistent with a large body of literature \cite{dongIntermodelSpreadPattern2020, armourSeasurfaceTemperaturePattern2024, andrewsDependenceRadiativeForcing2015,rugensteinDependenceGlobalRadiative2016}. These parameters do not enter the equilibrium calculations of ECS and $\phi$ (i.e., they primarily control the transient response, although $\epsilon$ has an effect in the calibration of the value of $\lambda_O$, see for instance \citeA{geoffroyTransientClimateResponse2013a}), and therefore we focus our discussion on the radiative and horizontal energy transport parameters.

\subsection{Two different pathways for the warming contrast}

A natural question is why $\phi_{eq}$ is relatively robust across models, as compared to the wide range of radiative parameters exhibited by CMIP6 models. To address this, it is useful to decompose the equilibrium values of $\Delta T_L$ and $\Delta T_O$ into their radiative and dynamical contributions. This provides a useful diagnostic for understanding how the warming contrast is achieved, and how different models achieve a similar warming contrast through different mechanisms. The equilibrium solution of Equation \ref{eq:emulator} can be written as:

\begin{subequations}
\label{eq:equilibrium}
\begin{align}
\Delta T_{L,4\times}
&=
\underbrace{\frac{F_L}{\lambda_L}}_{\text{Radiative}}
+
\underbrace{
\frac{F_H + \lambda_H \Delta T_{O,4\times}}
     {a_L\lambda_L}
}_{\text{Dynamical}},
\label{eq:equilibrium-land}
\\
\Delta T_{O,4\times}
&=
\underbrace{\frac{F_O}{\lambda_O}}_{\text{Radiative}}
-
\underbrace{
\frac{F_H + \lambda_H \Delta T_{O,4\times}}
     {a_O\lambda_O}
}_{\text{Dynamical}},
\label{eq:equilibrium-ocean}
\\
\Delta T_{D,4\times}
&=
\Delta T_{O,4\times}.
\label{eq:equilibrium-deep}
\end{align}

\end{subequations}
The first term on the right-hand side represents the radiative contribution, i.e. the equilibrium temperature that would arise if the atmospheric columns over land and ocean were dynamically uncoupled. The second term represents the dynamical energy exchange, which has opposite signs for the two columns. The resulting temperature response from the horizontal energy exchange is then modulated by the local radiative feedbacks, $\lambda_L$ and $\lambda_O$, and by the respective surface areas ($a_L$ and $a_O$).

\begin{figure}[bth]
    \centering
    \includegraphics[width=1.0\linewidth]{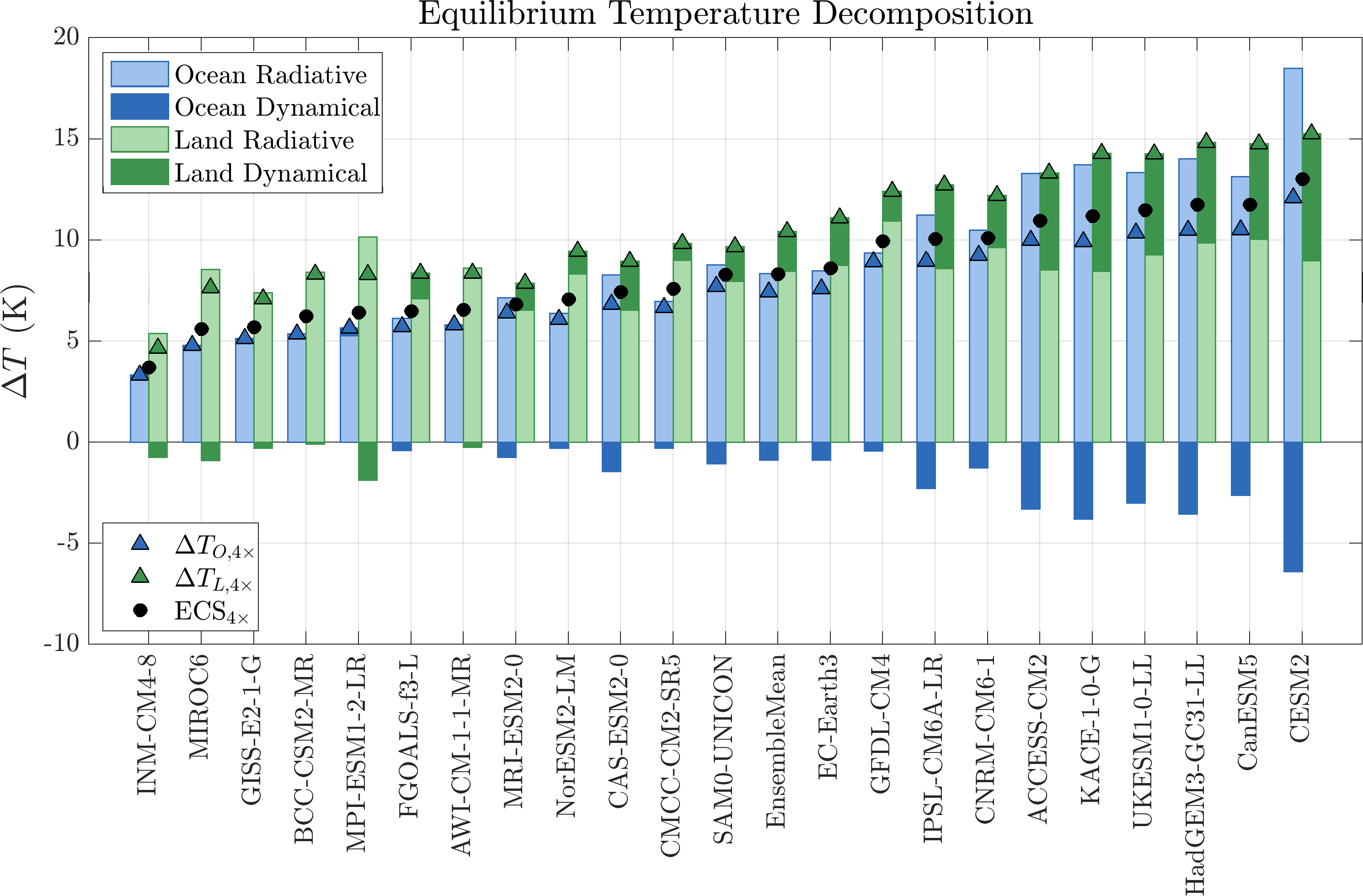}
    \caption{Decomposition of the equilibrium temperature response to a quadrupling of CO$_2$ over land and ocean into radiative and dynamical components, as defined in Equation \ref{eq:equilibrium}, for the 22 CMIP6 models that we consider in this work. }
    \label{fig:tempdecomposition}
\end{figure}

Figure \ref{fig:tempdecomposition} shows quantitative calculations of the radiative and dynamical terms based on the parameters in Table \ref{tab:params}. Models are ordered from left to right by increasing ECS.
As previously noted, the relative magnitude of the radiative contributions between land and ocean is not robust, i.e. roughly half of the CMIP6 models would warm the ocean more than land if the system was dynamically uncoupled, while the other half would have the opposite contrast. This reflects the model spread in the relative magnitudes of $\lambda_O$ and $\lambda_L$, as discussed in the previous section (i.e., only about half of the models have $\lambda_O > \lambda_L$).
Figure \ref{fig:tempdecomposition} adds an important additional insight: there is a clear relationship between high-ECS models and those with $\lambda_O < \lambda_L$. This arises because much of the intermodel spread is in $\lambda_O$, whereas $\lambda_L$ is more tightly constrained around $\sim$1 W m$^{-2}$ K$^{-1}$ and has a weaker influence on ECS. 
How do high-ECS models still produce a land-ocean warming contrast, given that their radiative feedbacks alone would favor greater ocean warming? The key lies in the dynamical feedback, which scales with $\Delta T_O$. In high-ECS models, the ocean warms substantially, which strengthens the dynamical transport of heat from ocean to land (approximately 0.17 PW K$^{-1}$). The strong ocean-to-land dynamical response overcomes the land-to-ocean fast adjustment $F_{H,4\times}$ (approximately 1 PW under 4×CO$_2$ forcing), strengthening the climatological heat transport from ocean to land. In Figure \ref{fig:tempdecomposition}, this manifests as a cooling tendency for the ocean for high-ECS models, and a warming tendency for low-ECS models. In other words, models with $\Delta T_{O,4\times} > 6$ K (corresponding roughly to ECS$_{2\times} > 3$ K) tend to exhibit a net heat transport increase from ocean to land at equilibrium. The strength of this effect increases with model ECS, as shown in Figure \ref{fig:tempdecomposition}.

This motivates a model classification into two broad regimes: (1) models with strongly stabilizing $\lambda_O$ (greater than approximately 1 W m$^{-2}$ K$^{-1}$), and (2) models with weaker, less stabilizing $\lambda_O$ (less than approximately 0.75 W m$^{-2}$ K$^{-1}$). The first regime corresponds to low-ECS models, in which the land-ocean warming contrast arises primarily from the radiative feedback contrast, and dynamics has a small heating tendency for the ocean. The second group corresponds to high-ECS models, in which radiative feedbacks alone would favor greater ocean warming than land. However, this does not occur because the dynamical feedback strengthens in high-ECS models and acts as a cooling tendency for the ocean. What emerges is that, regardless of the radiative configuration, MSE exchange robustly shifts the response toward $\phi > 1$, even when the radiative feedback configuration alone would favor greater ocean warming.  Physically, MSE exchange favors greater warming over land than over ocean ($\phi>1$) because the lower (and decreasing) relative humidity over land suppresses the increase in land MSE relative to temperature. At the same time, MSE exchange reduces the intermodel spread in $\phi$. The dry-static component of the transport response, associated with the exchange of $c_pT$, acts as a restoring mechanism: it preferentially promotes ocean warming in models with an unusually large land-ocean warming contrast (e.g., models where the radiative feedback over land is less stabilizing than over the ocean), and land warming in models with an unusually small contrast (e.g., models where the radiative feedback over land is more stabilizing than over the ocean). Thus, the MSE exchange both favors $\phi>1$ and constrains its intermodel variability.

\begin{figure}
    \centering
    \includegraphics[width=1.0\linewidth]{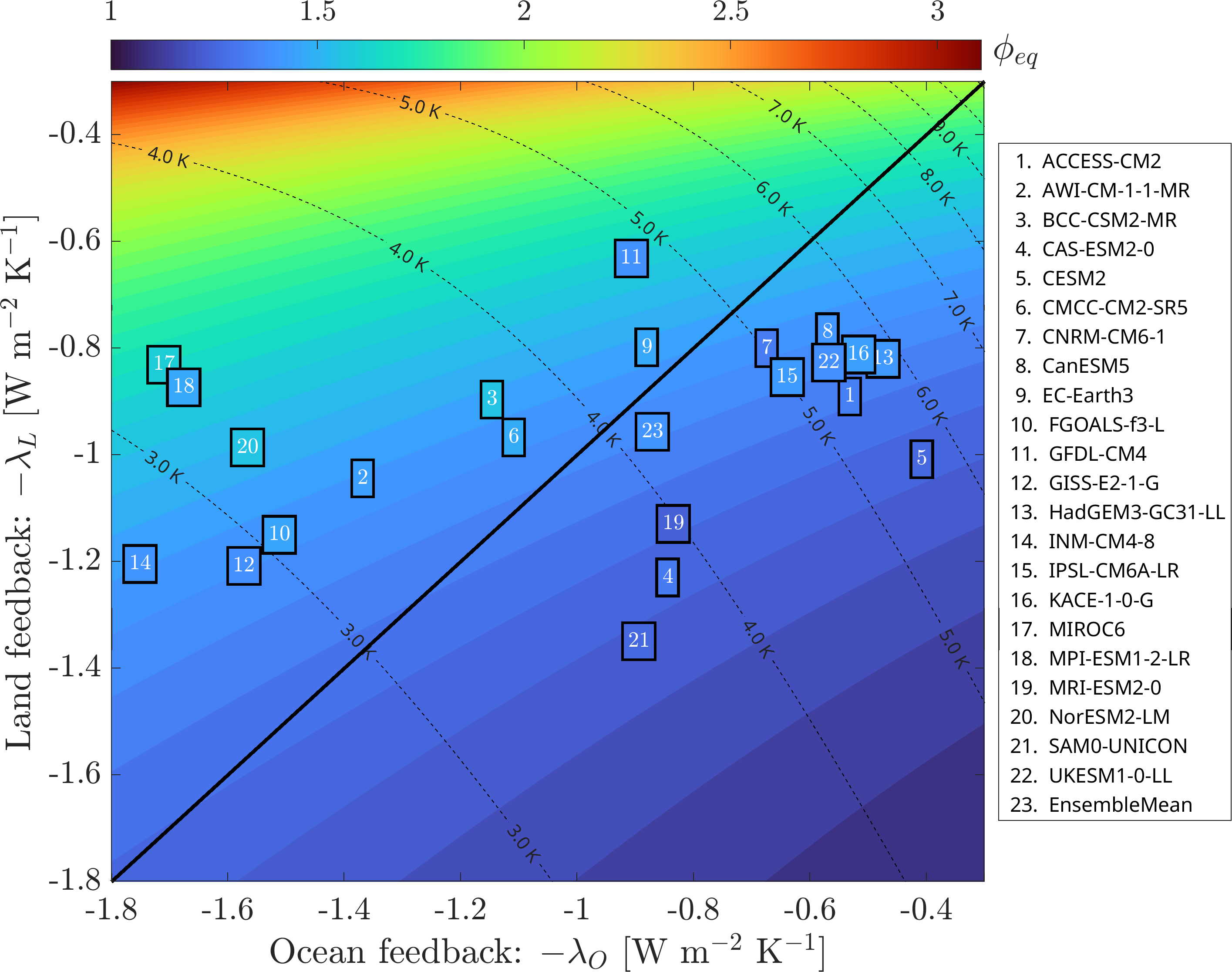}
    \caption{Equilibrium solution of Equation \ref{eq:emulator} for the land-ocean amplification factor $\phi$ (colored filled contours) and for the equilibrium climate sensitivity ECS$_{2\times}$ (dotted contours), as a function of the radiative feedbacks over land and ocean $\lambda_L$ and $\lambda_O$. The other parameters that play a role in $\phi$ and ECS$_{2\times}$, i.e. the forcing $F_{O,4\times}$, $F_{L,4\times}$ and the dynamical parameters $F_{H,4\times}$ and $\lambda_{H}$, are assumed to be correlated with $\lambda_O$ and $\lambda_L$ as detailed in Appendix C. Individual models are color-coded with their equilibrium $\phi$ (Table \ref{tab:params}), which could be different from the theoretical $\phi$ in the filled contours because of differences in $F_{O,4\times}$, $F_{L,4\times}$, $F_{H,4\times}$ and $\lambda_{H}$.}
    \label{fig:phase-space}
\end{figure}

Figure \ref{fig:phase-space} summarizes these results in the $\lambda_L$,$\lambda_O$ phase space and further illustrates the role of dynamical feedbacks. Colored filled contours show solutions of Equation \ref{eq:equilibrium} for $\phi_{eq}$, while dotted contours indicate solutions for ECS$_{2\times}$. Solutions are computed for each pair of $\lambda_L$ and $\lambda_O$, assuming that the remaining parameters ($F_H$, $F_L$, $F_O$, $\lambda_H$) covary with these two quantities (see Appendix C for details).
Values of $\phi_{eq}$ increase from the bottom-right corner (less stabilizing feedback over the ocean and more stabilizing feedback over land) toward the top-left corner (the opposite configuration). If land and ocean regions were dynamically uncoupled, any model that is in the phase space area below the 1:1 line would have $\phi < 1$ (roughly half of the CMIP6 models, e.g., CESM2, CanESM5, MRI-ESM2-0, assuming $F_L$=$F_O$). Interestingly, this does not happen even for the most favorable radiative configuration for $\phi < 1$ (the bottom right corner, where $\phi \sim 1.1$), because of two main reasons. First, a more stabilizing $\lambda_L$ is associated with a larger $\lambda_H$ (as the two are strongly correlated, as discussed above), leading to a stronger dynamical feedback \emph{rate} (i.e. energy transport per unit temperature change). Second, this configuration is associated with a high ECS (ECS$_{2\times} > 4$ K). The combination of a strong dynamical feedback rate and a warm model maintains the land-ocean warming contrast in the high-ECS regime. A final note is that models in Figure \ref{fig:phase-space} are color-coded by their $\phi_{eq}$, which does not exactly match the underlying contour values due to residual variability in $F_O$, $F_L$, $F_H$, and $\lambda_H$. In other words, the small mismatch reflects the fact that $\phi_{eq}$ is not fully determined by $\lambda_L$ and $\lambda_O$ alone (it would be if $F_O$, $F_L$, $F_H$, and $\lambda_H$ were perfectly correlated with $\lambda_L$ and $\lambda_O$, which is not the case). Nevertheless, knowing $\lambda_L$ and $\lambda_O$ and their correlation with $F_O$, $F_L$, $F_H$, and $\lambda_H$ explains much of the variability in $\phi_{eq}$ ($r=0.70$ between the true $\phi_{eq}$ and Figure \ref{fig:phase-space} approximation).

\subsection{The intermodel spread in $\phi_{eq}$}

What explains the intermodel spread in $\phi_{eq}$ shown in Table \ref{tab:params}? In this Section, we show that this spread is not fully explained by the parameters in the equal MSE framework, and that introducing radiative feedbacks helps explain a larger fraction of the intermodel differences in $\phi_{eq}$.

\begin{figure}
    \centering
    \includegraphics[width=1.0\linewidth]{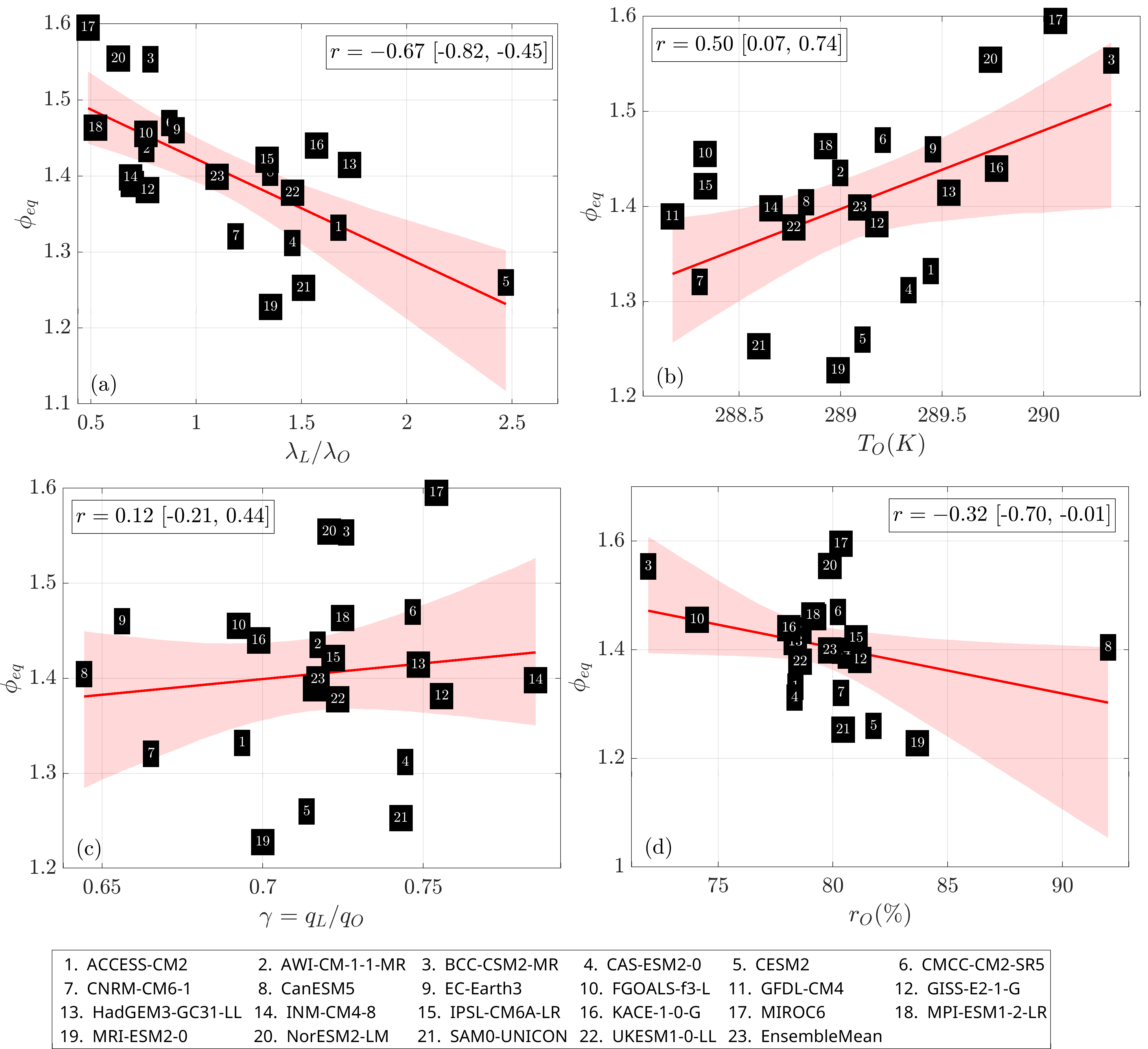}
    \caption{Correlation and best linear fit using ordinary least squares (red lines) between different potential predictors and the equilibrium land/ocean amplification factor $\phi_{eq}$: (a) ratio of radiative feedbacks over land and ocean $\lambda_L/\lambda_O$; (b) climatological ocean temperature $T_O$; (c) ratio of specific humidity over land and ocean $\gamma$ and (d) climatological relative humidity over the ocean $r_O$. $r_O$, $\gamma$ and $T_O$ are calculated from CMIP6 piControl experiments; $\lambda_L/\lambda_O$ and $\phi_{eq}$ are from Table \ref{tab:params}. The 95\% confidence intervals for the $r$ values and OLS regression lines are shown in square brackets and by red shading, respectively, and are estimated using bootstrapping.}
    \label{fig:correlations}
\end{figure}

According to the dynamical perspective (equal MSE framework), $\phi_{eq}$ only depends on $\gamma=q_L/q_O$, ocean relative humidity $r_O$, and the climatological ocean temperature $T_O$ through the temperature dependence of $q_{\text{sat}}$ ($q_O=r_Oq_{\text{sat}}(T_O)$), in addition to physical constants (Equation \ref{eq:phi-dynamical}). Specifically, it suggests that models with smaller $\gamma$, larger $r_O$, and warmer climatological oceans should exhibit larger values of $\phi_{eq}$. However, these relationships are only weakly reflected in the CMIP6 intermodel spread (Figure \ref{fig:correlations}). In particular, $\gamma$ varies substantially across models, spanning approximately 0.55--0.70 in piControl across models, but shows no significant correlation with $\phi_{eq}$. The lack of correlation may reflect limitations of the underlying assumptions in the dynamical perspective, such as the invariance of $\gamma$ under warming and the validity in the tropics. We have also tested an alternative definition of $q_O$ that is weighted by land fraction at each latitude as in \citeA{byrneTrendsContinentalTemperature2018}, but the correlation results are insensitive to that choice.
Similarly, $r_O$ only shows a weak correlation with $\phi_{eq}$ ($r=-0.32$), and the sign is opposite to that predicted by Equation \ref{eq:phi-dynamical}. This may partly reflect the limited intermodel spread in climatological ocean relative humidity, with most models clustered near $r_O\approx80\%$ (with a few exceptions), as well as differences in how near-surface relative humidity is diagnosed across models. Changes in $r_O$ under warming, which are neglected in the simple dynamical framework, may also contribute, together with changes in $\gamma$. There is some evidence for this interpretation: \citeA{geoffroyLandseaWarmingContrast2015} found a correlation between $\phi_{eq}$ and the ratio $k_O/k_L$ when $k_O$ and $k_L$ are calculated using the actual simulated changes in specific humidity, rather than assuming climate-invariant $r_O$ and $\gamma$. Finally, we find a weak positive relationship between climatological $T_O$ and $\phi_{eq}$ ($r$ = 0.50), i.e. models with warmer climatological oceans do tend to produce a larger equilibrium land-ocean warming contrast. This relationship is qualitatively consistent with the expectation from Equation \ref{eq:phi-dynamical} via the temperature dependence of the saturation specific humidity $q_{\text{sat}}$ in Equation \ref{eq:phi-dynamical}. Although the correlation is modest, it is potentially important because it links a climatological property of the model to its forced response, suggesting that $T_O$ may contain predictive information about $\phi_{eq}$.

A more predictive dimensionless number for $\phi_{eq}$ in CMIP6 models is $\lambda_L/\lambda_O$, which explains about half of the variance and shows a clear anticorrelation with CMIP6 $\phi_{eq}$ ($r=-0.67$, Figure \ref{fig:correlations}). This follows naturally from our description of the warming contrast mechanisms in the previous section, where Figure \ref{fig:phase-space} showed that models with low $\lambda_L/\lambda_O$ (i.e., in the top-left corner of the phase space) tend to have high $\phi_{eq}$ (because of a favorable radiative feedback configuration), and vice versa. Models with very high $\lambda_L/\lambda_O$ (e.g., CESM2) are typically high-ECS models (because of small $\lambda_O$), where a large $\phi$ is harder to maintain because of the radiative feedback configuration that favors more warming over the ocean than over land. This suggests a dependence of $\phi_{eq}$ on radiative feedbacks (and ECS) that has not been widely recognized before.

\subsection{The time-evolution of $\phi$}

Why does $\phi$ remain nearly invariant during historical and future scenario experiments, as well as in observations of recent climate, while evolving substantially with time in abrupt-4xCO$_2$ experiments? The transient evolution in abrupt-4xCO$_2$ experiments can be explained by the different heat capacities of land and ocean. It is less clear, however, why $\phi$ remains nearly time-invariant (or shows a very modest decrease) over the historical period and future scenarios. Several studies have investigated this weak temporal dependence of $\phi$ in 1pctCO$_2$ experiments, historical simulations, and observations \cite{boerRatioLandOcean2011, lambertRelationshipLandOcean2011, wallaceComparisonLandOcean2018}. Proposed explanations range from the nearly time-invariant structure of the leading spatial mode of temperature change \cite{boerRatioLandOcean2011}, to changes in atmospheric heat transport \cite{lambertRelationshipLandOcean2011}, and compensation between ocean heat uptake and atmospheric energy transport \cite{todaEnergyBudgetFramework2023}. Here, we show that the weak temporal variation of $\phi$ also follows naturally from Equation \ref{eq:emulator} and the timescales implied by its calibrated parameters, following arguments similar to \citeA{gianiOriginLimitsInvariant2026}.

In the following discussion, we approximate historical and future forcing as exponential, $F(t)\propto e^{rt}$, where $r$ is the forcing growth rate. Similar arguments apply to linear or polynomial forcing \cite{gianiOriginLimitsInvariant2026}. The solution to Equation \ref{eq:emulator} with exponential forcing can be written as:
\begin{equation}
\mathbf{\Delta T}(t)
=
\mathbf{A}e^{rt}
+
\mathbf{B}_1 e^{\mu_1 t}
+
\mathbf{B}_2 e^{\mu_2 t},
\end{equation}
where $\mathbf{\Delta T}(t)=[\Delta T_O(t),\Delta T_D(t)]$ and $\mu_1,\mu_2<0$ are the two eigenvalues describing the decaying transient modes. The response is therefore the sum of a forced component, $\mathbf{A}e^{rt}$, and two transient modes, $\mathbf{B}_1e^{\mu_1t}$ and $\mathbf{B}_2e^{\mu_2t}$, which decay from any initial imbalance. The land temperature $\Delta T_L$ is diagnostic and therefore does not introduce an additional mode. Once the transient modes become small relative to the forced response, $\Delta T_O$, $\Delta T_D$, and $\Delta T_L$ all evolve with the same time dependence, $\Delta T_j\propto e^{rt}$, and their ratios (including $\phi$) become approximately time-invariant. The relative importance of the slow transient mode, which decays with timescale $\tau$, compared with the forced mode scales as $e^{-t/\tau}/e^{rt}$. Thus, the e-folding timescale of the transient adjustment relative to the forced response is
\begin{equation}
\tau_{\rm rel}
=
\frac{1}{1/\tau_0+1/\tau},
\end{equation}
where $\tau_0=1/r$ is the forcing timescale.

For typical parameters in Equation \ref{eq:emulator} ($C_O=16$ W m$^{-2}$ K$^{-1}$ yr, $C_D=150$ W m$^{-2}$ K$^{-1}$ yr, $\lambda_O=1$ W m$^{-2}$ K$^{-1}$, and $\psi=1$ W m$^{-2}$ K$^{-1}$), the slowest eigenvalue is approximately $\mu_{\rm slow}=-0.004$ yr$^{-1}$, corresponding to an intrinsic adjustment timescale $\tau\approx250$ yr. Using a typical value of $\tau_0=50$ yr to mimic historical+SSP5-8.5 forcing trajectories \cite{gianiOriginLimitsInvariant2026}, this gives an e-folding timescale $\tau_{\rm rel}\approx40$ yr. A numerical evaluation of Equation \ref{eq:emulator} with the parameters above and forcing $F_i(t)=f_{i,4\times}\left[\exp(t/\tau_0)-1\right]$, where $f_{i,4\times}$ is chosen such that the forcing reaches $F_{i,4\times}$ over 350 years (1750--2100, conceptually similar to historical+SSP5-8.5), shows that $\phi$ becomes approximately constant after about $3\tau_{\rm rel}$, consistent with the argument above. In this simple numerical experiment, the decrease in $\phi$ between 1900 (approximately $3\tau_{\rm rel}$ after preindustrial) and 2100 is less than 0.01, consistent with the weak  temporal dependence noted in previous studies. This interpretation adds to the existing literature in that the time invariance of $\phi$ can be explained even in the absence of atmospheric heat transport, and originates from the decay of the intrinsic slow mode (dictated by the deep ocean) relative to the more rapidly evolving externally forced response.

\section{Discussion and Conclusions} 

We showed that the energetic and dynamical perspectives on the land-ocean warming contrast are linked through surface MSE exchange, and that different parameterizations of horizontal energy transport proposed in the literature are fundamentally equivalent. We summarize this connection in Equation \ref{eq:emulator}, which provides a minimal prognostic representation of the land-ocean warming contrast in climate models (an interpretable ``emulator"). Equation \ref{eq:emulator} reproduces the land-ocean warming contrast across CMIP6 models and provides a framework for interpreting the physical origin of their differences. Three main insights emerge from the interpretation of the emulated parameters:

\begin{enumerate}

   \item Physically, the tendency of models toward $\phi>1$ is rooted in the lower relative humidity over land than over the ocean, which promotes greater land warming. This can be understood either through atmospheric MSE constraints \cite{byrneTrendsContinentalTemperature2018} or from energy balance \cite{suttonLandSeaWarming2007}, which we show are consistent. However, we show that the value of $\phi$ in any given model is also strongly influenced by the land--ocean radiative feedback contrast, which determines whether radiative processes alone would favor greater warming over land or over the ocean. The exchange of MSE is a restoring mechanism acting on this radiative baseline: if the radiative feedback configuration favors unusually strong land warming, atmospheric energy transport shifts energy toward the ocean and reduces the contrast; conversely, if radiative feedbacks favor stronger ocean warming, MSE transport shifts energy toward land. In both cases, the restoring tendency is toward a state with $\phi>1$, owing to the lower (and decreasing) relative humidity over land. Thus, MSE exchange not only helps maintain $\phi>1$, but also dampens the intermodel spread in $\phi$ generated by differences in radiative feedbacks.
    
    \item Because of the influence of the radiative feedback configuration, and the large intermodel spread in $\lambda_O$, the intermodel spread in $\phi_{eq}$ (although moderated by the restoring effect of MSE exchange) is closely related to the ratio $\lambda_L/\lambda_O$. In contrast, the land-ocean specific-humidity ratio $\gamma$ and ocean relative humidity $r_O$ play comparatively smaller roles. Somewhat surprisingly, we also find a correlation between climatological ocean temperature $T_O$ and $\phi_{eq}$ across CMIP6 models, which we tentatively attribute to the temperature dependence of saturation specific humidity through the Clausius--Clapeyron relationship. Further work is needed to better understand this connection.

    \item The negative correlation between ECS and $\phi_{eq}$, noted by \citeA{todaEnergyBudgetFramework2021}, can be understood through the dependence of $\phi_{eq}$ on the radiative feedback ratio $\lambda_L/\lambda_O$. High-ECS models tend to have a small stabilizing effective feedback parameter over the ocean (small $\lambda_O$), which favors strong ocean warming and a large $\lambda_L/\lambda_O$. Because $\phi_{eq}$ is anticorrelated with this ratio, these models tend to exhibit a smaller land--ocean warming contrast. In contrast, low-ECS models tend to have more stabilizing ocean feedbacks (large $\lambda_O$), which suppress ocean warming and favor a larger $\phi_{eq}$. Thus, differences in the land--ocean radiative feedback configuration help explain why ECS and $\phi_{eq}$ are moderately anticorrelated across models.
    \end{enumerate}

Beyond the land-ocean warming contrast, similar emulator approaches could be applied to other problems in climate dynamics, provided that a suitable physical framework can serve as the basis for the emulator. This would enhance the interpretability of CMIP6 output (which involves many interacting parameterizations and processes) by clarifying both the mechanisms that shape the spatial patterns of response and the sources of intermodel spread. This approach could also be extended to include multiple model configurations within each modeling center, providing an efficient diagnostic of how different configurations of a given model (e.g., CESM2 vs. CESM2-WACCM6) affect key properties of the simulated climate system.

For impact assessment, this framework could be incorporated into the IPCC ``emulator" approach, which uses simple climate models such as FaIR \cite{millarModifiedImpulseresponseRepresentation2017,smithFaircalibrateV141Calibration2024} and MAGICC \cite{meinshausenEmulatingCoupledAtmosphereocean2011} to efficiently explore global temperature responses under different emission scenarios. Equation \ref{eq:emulator} provides additional granularity (land/ocean temperatures, radiative fluxes, and changes in heat transport), while retaining the same computational efficiency. For emulators  which focus on land impacts, such as MESMER \cite{beuschEmulatingEarthSystem2020,bauerMESMERV100Consolidating2025}, Equation \ref{eq:emulator} offers a minimal representation of the land-ocean warming contrast that directly enables an estimate of the land temperature response.

Finally, a natural extension of this work would be to constrain some of the parameters in Equation \ref{eq:emulator} directly using observations of past and present-day climate. Despite the relatively robust response of the land-ocean amplification factor, models differ substantially in their radiative responses  over land and ocean (and their relative magnitude), as well as in changes in land-ocean heat transport. As global mean temperatures increase, the forced signal in observations of such quantities should become more distinguishable from internal variability, potentially offering an avenue to start ruling out some of the intermodel spread.

\appendix
\section{CMIP6 data}
 
We use CMIP6 models that provide monthly-averaged output for the following variables in the piControl, 1pctCO2, and abrupt-4xCO2 experiments: tas, rsut, rlut, rsdt, rsus, rsds, rlus, rlds, hfss, hfls, hurs, and huss. These standard CMIP6 variables correspond to surface air temperature, specific humidity, relative humidity, as well as the radiative and turbulent components of surface and top-of-atmosphere (TOA) fluxes. piControl is a preindustrial unforced experiment, whereas abrupt-4xCO2 and 1pctCO2 are forced experiments with an abrupt change to 4 times the CO$_2$ concentration or a 1\% increase in CO$_2$ concentration per year, respectively. We only include  models for which these variables were available at the time of data processing and that provide at least 150 years of simulations for each of the three experiments. When multiple models are available from a single modeling center within CMIP6, we retain only one configuration per modeling center. This selection yields a total of 22 CMIP6 models.

\section{Calibration and validation of the emulator}

We use 150 years of monthly data from abrupt-4xCO2 experiments to tune the 10 parameters of Equation \ref{eq:emulator}. If multiple realizations are present within a model, we first average the realizations to obtain an ensemble average. We first average monthly data to annual averages, and calculate anomalies with respect to piControl climatology. TOA radiative parameters for land and ocean ($F_O$, $\lambda_O$, $F_L$, $\lambda_L$) are calculated by linearly regressing the total TOA flux anomalies against their respective temperatures. First guesses of $F_H$ and $\lambda_H$ are obtained by regressing the net atmospheric heating term over the ocean (net TOA flux - net surface flux) against $\Delta T_O$. First guesses of the thermal inertia parameters, $\psi$, $C_O$ and $C_D$ and the ocean heat uptake efficacy $\epsilon$ are calculated with a similar procedure to \citeA{geoffroyTransientClimateResponse2013a,geoffroyTransientClimateResponse2013}, which exploits known analytical solutions of the transient two-box energy balance model for early and late times of abrupt-4xCO2 to calculate fast and slow time scales. Estimates of $F_H$, $\lambda_H$, $\psi$, $C_O$ and $C_D$ and $\epsilon$ are then refined through an optimization routine to minimize the error between emulated and simulated temperatures. We use the interior point method implemented in MATLAB's \texttt{fmincon} \cite{byrdTrustRegionMethod2000} for the optimization, starting from plausible values for $F_H$ and $\lambda_H$ from the linear regression and imposing physical boundaries on the parameters to restrict the search space (e.g., heat capacities and $\psi$ greater than zero). Results are almost insensitive to different optimization choices (not shown). We use 1pctCO2 experiments for validation. The forcing in 1pctCO2 for Equation \ref{eq:emulator} is prescribed as $F_i(t) = F_{i,4\times} R( [C_0] 1.01^t)/R(4[C_0])$, where $i \in \{L, O, H\}$  and $R(\cdot)$ is the empirical function that maps CO$_2$ concentrations to effective radiative forcing, derived from line-by-line calculations in \citeA{meinshausenSharedSocioeconomicPathway2020}, $[C_0]$ is 277 ppm, and $t$ is in years. Changes in heat transport in CMIP6 models (e.g., the bottom row in Figure \ref{fig:validation}) are inferred from changes in net atmospheric heating over land and ocean (i.e., the difference between net TOA and surface fluxes), under the assumption that any non-zero imbalance within the atmospheric column must be balanced by atmospheric dynamics due to the small heat capacity in the atmosphere. This approximation is not exact, as we neglect minor terms in the surface energy budget (i.e., contributions other than radiative, sensible, and latent heat fluxes), and because climate models may exhibit small residual energy imbalances. We find and subtract the small residual imbalance in $\Delta H$ for each model, which is typically smaller than 0.05 PW.

\section{Correlations between different emulator parameters}
We calculate the best linear fit across parameters in Equation \ref{eq:emulator} to reduce the dimensionality of the equilibrium problem to two parameters ($\lambda_L$ and $\lambda_O$) in Figure \ref{fig:phase-space}. Correlations are calculated using all 22 CMIP6 models that we consider. In other words, we fit a simple linear model $y = \alpha_0 + \alpha_1 \lambda_L + \alpha_2 \lambda_O$, where $y$ represents either $F_L$, $F_H$, $F_O$ or $\lambda_H$ and $\alpha_i$ are the linear regression best fit coefficients. We expect a high degree of correlation between $\lambda_L$ and $\lambda_H$ on the basis of Equation \ref{eq:lh}, and indeed the simple linear model based on $\lambda_L$ and $\lambda_O$ explains 87\% of the intermodel variability in $\lambda_H$ (mostly from $\lambda_L$). The forcing parameters are less correlated with $\lambda_L$ and $\lambda_O$, which is what introduces the variability in Figure \ref{fig:phase-space} between the model $\phi$ and the theoretical calculations in the contours. Specifically, a linear model based on $\lambda_L$ and $\lambda_O$ only explains about 16\%, 30\% and 18\% of the total variability in $F_O$, $F_L$ and $F_H$, respectively.

%

%



%
%

\section*{Open Research Section}
All CMIP6 data used in this study are publicly available through the Earth System Grid Federation (ESGF). The specific models, experiments, and variables used are documented in Appendix A. Dataset-specific DOIs are provided through the ESGF CMIP6 archive.
\section*{Conflict of Interest declaration}
The authors declare there are no conflicts of interest for this manuscript.

\acknowledgments
We acknowledge support from Schmidt Sciences and the MIT Climate Grand Challenges through the Bringing Computation to the Climate Challenge (BC3) project. We also acknowledge the MIT Center for Sustainability Science and Strategy for computing resources through the svante cluster. We are grateful to the BC3 team for insightful discussions about this work.

%
%

\bibliography{references_cited_only}

@article{byrdTrustRegionMethod2000,
	title = {A trust region method based on interior point techniques for nonlinear programming},
	volume = {89},
	copyright = {http://www.springer.com/tdm},
	issn = {0025-5610},
	url = {http://link.springer.com/10.1007/PL00011391},
	doi = {10.1007/PL00011391},
	language = {en},
	number = {1},
	journal = {Mathematical Programming},
	author = {Byrd, Richard H. and Gilbert, Jean Charles and Nocedal, Jorge},
	month = nov,
	year = {2000},
	pages = {149--185},
}

@article{tebaldiEmulatorsClimateModel2025a,
	title = {Emulators of {Climate} {Model} {Output}},
	volume = {50},
	copyright = {http://creativecommons.org/licenses/by/4.0/},
	issn = {1543-5938, 1545-2050},
	url = {https://www.annualreviews.org/content/journals/10.1146/annurev-environ-012125-085838},
	doi = {10.1146/annurev-environ-012125-085838},
	language = {en},
	number = {1},
	journal = {Annual Review of Environment and Resources},
	author = {Tebaldi, C. and Selin, N.E. and Ferrari, R. and Flierl, G.},
	month = oct,
	year = {2025},
	pages = {709--737},
}

@misc{bauerMESMERV100Consolidating2025,
	title = {{MESMER} v1.0.0: {Consolidating} the {Modular} {Earth} {System} {Model} {Emulator} into a {Sustainable} {Research} {Software} {Package}},
	copyright = {https://creativecommons.org/licenses/by/4.0/},
	shorttitle = {{MESMER} v1.0.0},
	url = {https://egusphere.copernicus.org/preprints/2025/egusphere-2025-4917/},
	doi = {10.5194/egusphere-2025-4917},
	publisher = {Climate and Earth system modeling},
	author = {Bauer, Victoria M. and Hauser, Mathias and Quilcaille, Yann and Schöngart, Sarah and Gudmundsson, Lukas and Seneviratne, Sonia I.},
	month = dec,
	year = {2025},
}

@article{meinshausenEmulatingCoupledAtmosphereocean2011,
	title = {Emulating coupled atmosphere-ocean and carbon cycle models with a simpler model, {MAGICC6} – {Part} 1: {Model} description and calibration},
	volume = {11},
	copyright = {https://creativecommons.org/licenses/by/3.0/},
	issn = {1680-7324},
	shorttitle = {Emulating coupled atmosphere-ocean and carbon cycle models with a simpler model, {MAGICC6} – {Part} 1},
	url = {https://acp.copernicus.org/articles/11/1417/2011/},
	doi = {10.5194/acp-11-1417-2011},
	language = {en},
	number = {4},
	journal = {Atmospheric Chemistry and Physics},
	author = {Meinshausen, M. and Raper, S. C. B. and Wigley, T. M. L.},
	month = feb,
	year = {2011},
	pages = {1417--1456},
}

@article{smithFaircalibrateV141Calibration2024,
	title = {fair-calibrate v1.4.1: calibration, constraining, and validation of the {FaIR} simple climate model for reliable future climate projections},
	volume = {17},
	copyright = {https://creativecommons.org/licenses/by/4.0/},
	issn = {1991-9603},
	shorttitle = {fair-calibrate v1.4.1},
	url = {https://gmd.copernicus.org/articles/17/8569/2024/},
	doi = {10.5194/gmd-17-8569-2024},
	language = {en},
	number = {23},
	journal = {Geoscientific Model Development},
	author = {Smith, Chris and Cummins, Donald P. and Fredriksen, Hege-Beate and Nicholls, Zebedee and Meinshausen, Malte and Allen, Myles and Jenkins, Stuart and Leach, Nicholas and Mathison, Camilla and Partanen, Antti-Ilari},
	month = dec,
	year = {2024},
	pages = {8569--8592},
}

@article{boerRatioLandOcean2011,
	title = {The ratio of land to ocean temperature change under global warming},
	volume = {37},
	copyright = {http://www.springer.com/tdm},
	issn = {0930-7575, 1432-0894},
	url = {http://link.springer.com/10.1007/s00382-011-1112-3},
	doi = {10.1007/s00382-011-1112-3},
	language = {en},
	number = {11-12},
	journal = {Climate Dynamics},
	author = {Boer, G. J.},
	month = dec,
	year = {2011},
	pages = {2253--2270},
}

@article{wallaceComparisonLandOcean2018,
	title = {Comparison of land–ocean warming ratios in updated observed records and {CMIP5} climate models},
	volume = {13},
	issn = {1748-9326},
	url = {https://iopscience.iop.org/article/10.1088/1748-9326/aae46f},
	doi = {10.1088/1748-9326/aae46f},
	language = {en},
	number = {11},
	journal = {Environmental Research Letters},
	author = {Wallace, C J and Joshi, M},
	month = nov,
	year = {2018},
	pages = {114011},
}

@article{gianiOriginLimitsInvariant2026,
	title = {Origin and {Limits} of {Invariant} {Warming} {Patterns} in {Climate} {Models}},
	volume = {39},
	copyright = {http://www.ametsoc.org/PUBSReuseLicenses},
	issn = {0894-8755, 1520-0442},
	url = {https://journals.ametsoc.org/view/journals/clim/39/7/JCLI-D-24-0683.1.xml},
	doi = {10.1175/JCLI-D-24-0683.1},
	number = {7},
	journal = {Journal of Climate},
	author = {Giani, Paolo and Fiore, Arlene M. and Flierl, Glenn and Ferrari, Raffaele and Selin, Noelle E.},
	month = apr,
	year = {2026},
	pages = {1681--1701},
}

@article{brethertonInsightsLowlatitudeCloud2015,
	title = {Insights into low-latitude cloud feedbacks from high-resolution models},
	volume = {373},
	issn = {1364-503X, 1471-2962},
	url = {https://royalsocietypublishing.org/doi/10.1098/rsta.2014.0415},
	doi = {10.1098/rsta.2014.0415},
	language = {en},
	number = {2054},
	journal = {Philosophical Transactions of the Royal Society A: Mathematical, Physical and Engineering Sciences},
	author = {Bretherton, Christopher S.},
	month = nov,
	year = {2015},
	pages = {20140415},
}

@article{seltzerTerrestrialAmplificationPresent2023,
	title = {Terrestrial amplification of past, present, and future climate change},
	volume = {9},
	issn = {2375-2548},
	url = {https://www.science.org/doi/10.1126/sciadv.adf8119},
	doi = {10.1126/sciadv.adf8119},
	language = {en},
	number = {6},
	journal = {Science Advances},
	author = {Seltzer, Alan M. and Blard, Pierre-Henri and Sherwood, Steven C. and Kageyama, Masa},
	month = feb,
	year = {2023},
	pages = {eadf8119},
}

@article{geoffroyLandseaWarmingContrast2015,
	title = {Land-sea warming contrast: the role of the horizontal energy transport},
	volume = {45},
	issn = {0930-7575, 1432-0894},
	shorttitle = {Land-sea warming contrast},
	url = {http://link.springer.com/10.1007/s00382-015-2552-y},
	doi = {10.1007/s00382-015-2552-y},
	language = {en},
	number = {11-12},
	journal = {Climate Dynamics},
	author = {Geoffroy, Olivier and Saint-Martin, David and Voldoire, Aurore},
	month = dec,
	year = {2015},
	pages = {3493--3511},
}

@article{dommengetOceansRoleContinental2009,
	title = {The {Ocean}’s {Role} in {Continental} {Climate} {Variability} and {Change}},
	volume = {22},
	issn = {1520-0442, 0894-8755},
	url = {http://journals.ametsoc.org/doi/10.1175/2009JCLI2778.1},
	doi = {10.1175/2009jcli2778.1},
	language = {en},
	number = {18},
	journal = {Journal of Climate},
	publisher = {American Meteorological Society},
	author = {Dommenget, Dietmar},
	month = sep,
	year = {2009},
	pages = {4939--4952},
}

@article{byrneUnderstandingDecreasesLand2016,
	title = {Understanding {Decreases} in {Land} {Relative} {Humidity} with {Global} {Warming}: {Conceptual} {Model} and {GCM} {Simulations}},
	volume = {29},
	issn = {0894-8755, 1520-0442},
	shorttitle = {Understanding {Decreases} in {Land} {Relative} {Humidity} with {Global} {Warming}},
	url = {http://journals.ametsoc.org/doi/10.1175/JCLI-D-16-0351.1},
	doi = {10.1175/JCLI-D-16-0351.1},
	language = {en},
	number = {24},
	journal = {Journal of Climate},
	author = {Byrne, Michael P. and O’Gorman, Paul A.},
	month = dec,
	year = {2016},
	pages = {9045--9061},
}

@article{lambertControlLandoceanTemperature2007,
	title = {Control of land‐ocean temperature contrast by ocean heat uptake},
	volume = {34},
	copyright = {http://onlinelibrary.wiley.com/termsAndConditions\#vor},
	issn = {0094-8276, 1944-8007},
	url = {https://agupubs.onlinelibrary.wiley.com/doi/10.1029/2007GL029755},
	doi = {10.1029/2007GL029755},
	language = {en},
	number = {13},
	journal = {Geophysical Research Letters},
	author = {Lambert, F. Hugo and Chiang, John C. H.},
	month = jul,
	year = {2007},
	pages = {2007GL029755},
}

@article{lambertRelationshipLandOcean2011,
	title = {The {Relationship} between {Land}–{Ocean} {Surface} {Temperature} {Contrast} and {Radiative} {Forcing}},
	volume = {24},
	issn = {0894-8755, 1520-0442},
	url = {http://journals.ametsoc.org/doi/10.1175/2011JCLI3893.1},
	doi = {10.1175/2011JCLI3893.1},
	language = {en},
	number = {13},
	journal = {Journal of Climate},
	author = {Lambert, F. Hugo and Webb, Mark J. and Joshi, Manoj M.},
	month = jul,
	year = {2011},
	pages = {3239--3256},
}

@article{todaEnergyBudgetFramework2023,
	title = {An {Energy} {Budget} {Framework} to {Understand} {Mechanisms} of {Land}–{Ocean} {Warming} {Contrast} {Induced} by {Increasing} {Greenhouse} {Gases}. {Part} {II}: {Transient} {Climate} {State}},
	volume = {36},
	copyright = {http://www.ametsoc.org/PUBSReuseLicenses},
	issn = {0894-8755, 1520-0442},
	shorttitle = {An {Energy} {Budget} {Framework} to {Understand} {Mechanisms} of {Land}–{Ocean} {Warming} {Contrast} {Induced} by {Increasing} {Greenhouse} {Gases}. {Part} {II}},
	url = {https://journals.ametsoc.org/view/journals/clim/36/13/JCLI-D-22-0483.1.xml},
	doi = {10.1175/JCLI-D-22-0483.1},
	language = {en},
	number = {13},
	journal = {Journal of Climate},
	author = {Toda, Masaki and Yoshimori, Masakazu and Watanabe, Masahiro},
	month = jul,
	year = {2023},
	pages = {4307--4326},
}

@article{wintonImportanceOceanHeat2010,
	title = {Importance of {Ocean} {Heat} {Uptake} {Efficacy} to {Transient} {Climate} {Change}},
	volume = {23},
	issn = {1520-0442, 0894-8755},
	url = {http://journals.ametsoc.org/doi/10.1175/2009JCLI3139.1},
	doi = {10.1175/2009JCLI3139.1},
	language = {en},
	number = {9},
	journal = {Journal of Climate},
	author = {Winton, Michael and Takahashi, Ken and Held, Isaac M.},
	month = may,
	year = {2010},
	pages = {2333--2344},
}

@article{geoffroyTransientClimateResponse2013a,
	title = {Transient {Climate} {Response} in a {Two}-{Layer} {Energy}-{Balance} {Model}. {Part} {II}: {Representation} of the {Efficacy} of {Deep}-{Ocean} {Heat} {Uptake} and {Validation} for {CMIP5} {AOGCMs}},
	volume = {26},
	issn = {0894-8755, 1520-0442},
	shorttitle = {Transient {Climate} {Response} in a {Two}-{Layer} {Energy}-{Balance} {Model}. {Part} {II}},
	url = {https://journals.ametsoc.org/view/journals/clim/26/6/jcli-d-12-00196.1.xml},
	doi = {10.1175/JCLI-D-12-00196.1},
	language = {en},
	number = {6},
	journal = {Journal of Climate},
	author = {Geoffroy, O. and Saint-Martin, D. and Bellon, G. and Voldoire, A. and Olivié, D. J. L. and Tytéca, S.},
	month = mar,
	year = {2013},
	pages = {1859--1876},
}

@article{meinshausenSharedSocioeconomicPathway2020,
	title = {The shared socio-economic pathway ({SSP}) greenhouse gas concentrations and their extensions to 2500},
	volume = {13},
	copyright = {https://creativecommons.org/licenses/by/4.0/},
	issn = {1991-9603},
	url = {https://gmd.copernicus.org/articles/13/3571/2020/},
	doi = {10.5194/gmd-13-3571-2020},
	language = {en},
	number = {8},
	journal = {Geoscientific Model Development},
	author = {Meinshausen, Malte and Nicholls, Zebedee R. J. and Lewis, Jared and Gidden, Matthew J. and Vogel, Elisabeth and Freund, Mandy and Beyerle, Urs and Gessner, Claudia and Nauels, Alexander and Bauer, Nico and Canadell, Josep G. and Daniel, John S. and John, Andrew and Krummel, Paul B. and Luderer, Gunnar and Meinshausen, Nicolai and Montzka, Stephen A. and Rayner, Peter J. and Reimann, Stefan and Smith, Steven J. and Van Den Berg, Marten and Velders, Guus J. M. and Vollmer, Martin K. and Wang, Ray H. J.},
	month = aug,
	year = {2020},
	pages = {3571--3605},
}

@article{geoffroyTransientClimateResponse2013,
	title = {Transient {Climate} {Response} in a {Two}-{Layer} {Energy}-{Balance} {Model}. {Part} {I}: {Analytical} {Solution} and {Parameter} {Calibration} {Using} {CMIP5} {AOGCM} {Experiments}},
	volume = {26},
	issn = {0894-8755, 1520-0442},
	shorttitle = {Transient {Climate} {Response} in a {Two}-{Layer} {Energy}-{Balance} {Model}. {Part} {I}},
	url = {https://journals.ametsoc.org/view/journals/clim/26/6/jcli-d-12-00195.1.xml},
	doi = {10.1175/JCLI-D-12-00195.1},
	language = {en},
	number = {6},
	journal = {Journal of Climate},
	author = {Geoffroy, O. and Saint-Martin, D. and Olivié, D. J. L. and Voldoire, A. and Bellon, G. and Tytéca, S.},
	month = mar,
	year = {2013},
	pages = {1841--1857},
}

@article{todaEnergyBudgetFramework2021,
	title = {An energy budget framework to understand mechanisms of land–ocean warming contrast induced by increasing greenhouse gases {Part} {I}: {Near}-equilibrium state},
	issn = {0894-8755, 1520-0442},
	shorttitle = {An energy budget framework to understand mechanisms of land–ocean warming contrast induced by increasing greenhouse gases {Part} {I}},
	url = {https://journals.ametsoc.org/view/journals/clim/aop/JCLI-D-21-0302.1/JCLI-D-21-0302.1.xml},
	doi = {10.1175/JCLI-D-21-0302.1},
	language = {en},
	journal = {Journal of Climate},
	author = {Toda, Masaki and Watanabe, Masahiro and Yoshimori, Masakazu},
	month = sep,
	year = {2021},
	pages = {1--63},
}

@article{byrneLandOceanWarming2013,
	title = {Land–{Ocean} {Warming} {Contrast} over a {Wide} {Range} of {Climates}: {Convective} {Quasi}-{Equilibrium} {Theory} and {Idealized} {Simulations}},
	volume = {26},
	issn = {0894-8755, 1520-0442},
	shorttitle = {Land–{Ocean} {Warming} {Contrast} over a {Wide} {Range} of {Climates}},
	url = {http://journals.ametsoc.org/doi/10.1175/JCLI-D-12-00262.1},
	doi = {10.1175/JCLI-D-12-00262.1},
	language = {en},
	number = {12},
	journal = {Journal of Climate},
	author = {Byrne, Michael P. and O’Gorman, Paul A.},
	month = jun,
	year = {2013},
	pages = {4000--4016},
}

@article{dongIntermodelSpreadPattern2020,
	title = {Intermodel {Spread} in the {Pattern} {Effect} and {Its} {Contribution} to {Climate} {Sensitivity} in {CMIP5} and {CMIP6} {Models}},
	volume = {33},
	issn = {0894-8755, 1520-0442},
	url = {https://journals.ametsoc.org/view/journals/clim/33/18/jcliD191011.xml},
	doi = {10.1175/JCLI-D-19-1011.1},
	language = {en},
	number = {18},
	journal = {Journal of Climate},
	author = {Dong, Yue and Armour, Kyle C. and Zelinka, Mark D. and Proistosescu, Cristian and Battisti, David S. and Zhou, Chen and Andrews, Timothy},
	month = sep,
	year = {2020},
	pages = {7755--7775},
}

@article{sherwoodAssessmentEarthClimate2020,
	title = {An {Assessment} of {Earth}'s {Climate} {Sensitivity} {Using} {Multiple} {Lines} of {Evidence}},
	volume = {58},
	issn = {8755-1209, 1944-9208},
	url = {https://agupubs.onlinelibrary.wiley.com/doi/10.1029/2019RG000678},
	doi = {10.1029/2019RG000678},
	language = {en},
	number = {4},
	journal = {Reviews of Geophysics},
	author = {Sherwood, S. C. and Webb, M. J. and Annan, J. D. and Armour, K. C. and Forster, P. M. and Hargreaves, J. C. and Hegerl, G. and Klein, S. A. and Marvel, K. D. and Rohling, E. J. and Watanabe, M. and Andrews, T. and Braconnot, P. and Bretherton, C. S. and Foster, G. L. and Hausfather, Z. and Von Der Heydt, A. S. and Knutti, R. and Mauritsen, T. and Norris, J. R. and Proistosescu, C. and Rugenstein, M. and Schmidt, G. A. and Tokarska, K. B. and Zelinka, M. D.},
	month = dec,
	year = {2020},
	pages = {e2019RG000678},
}

@article{suttonLandSeaWarming2007,
	title = {Land/sea warming ratio in response to climate change: {IPCC} {AR4} model results and comparison with observations},
	volume = {34},
	copyright = {http://onlinelibrary.wiley.com/termsAndConditions\#vor},
	issn = {0094-8276, 1944-8007},
	shorttitle = {Land/sea warming ratio in response to climate change},
	url = {https://agupubs.onlinelibrary.wiley.com/doi/10.1029/2006GL028164},
	doi = {10.1029/2006GL028164},
	language = {en},
	number = {2},
	journal = {Geophysical Research Letters},
	author = {Sutton, Rowan T. and Dong, Buwen and Gregory, Jonathan M.},
	month = jan,
	year = {2007},
	pages = {2006GL028164},
}

@article{joshiMechanismsLandSea2008,
	title = {Mechanisms for the land/sea warming contrast exhibited by simulations of climate change},
	volume = {30},
	copyright = {http://www.springer.com/tdm},
	issn = {0930-7575, 1432-0894},
	url = {http://link.springer.com/10.1007/s00382-007-0306-1},
	doi = {10.1007/s00382-007-0306-1},
	language = {en},
	number = {5},
	journal = {Climate Dynamics},
	author = {Joshi, Manoj M. and Gregory, Jonathan M. and Webb, Mark J. and Sexton, David M. H. and Johns, Tim C.},
	month = apr,
	year = {2008},
	pages = {455--465},
}

@article{gregoryNewMethodDiagnosing2004,
	title = {A new method for diagnosing radiative forcing and climate sensitivity},
	volume = {31},
	copyright = {http://onlinelibrary.wiley.com/termsAndConditions\#vor},
	issn = {0094-8276, 1944-8007},
	url = {https://agupubs.onlinelibrary.wiley.com/doi/10.1029/2003GL018747},
	doi = {10.1029/2003GL018747},
	language = {en},
	number = {3},
	journal = {Geophysical Research Letters},
	author = {Gregory, J. M. and Ingram, W. J. and Palmer, M. A. and Jones, G. S. and Stott, P. A. and Thorpe, R. B. and Lowe, J. A. and Johns, T. C. and Williams, K. D.},
	month = feb,
	year = {2004},
	pages = {2003GL018747},
}

@article{bonyCloudsCirculationClimate2015,
	title = {Clouds, circulation and climate sensitivity},
	volume = {8},
	issn = {1752-0894, 1752-0908},
	url = {https://www.nature.com/articles/ngeo2398},
	doi = {10.1038/ngeo2398},
	language = {en},
	number = {4},
	journal = {Nature Geoscience},
	author = {Bony, Sandrine and Stevens, Bjorn and Frierson, Dargan M. W. and Jakob, Christian and Kageyama, Masa and Pincus, Robert and Shepherd, Theodore G. and Sherwood, Steven C. and Siebesma, A. Pier and Sobel, Adam H. and Watanabe, Masahiro and Webb, Mark J.},
	month = apr,
	year = {2015},
	pages = {261--268},
}

@article{byrneTrendsContinentalTemperature2018,
	title = {Trends in continental temperature and humidity directly linked to ocean warming},
	volume = {115},
	issn = {0027-8424, 1091-6490},
	url = {https://pnas.org/doi/full/10.1073/pnas.1722312115},
	doi = {10.1073/pnas.1722312115},
	language = {en},
	number = {19},
	journal = {Proceedings of the National Academy of Sciences},
	author = {Byrne, Michael P. and O’Gorman, Paul A.},
	month = may,
	year = {2018},
	pages = {4863--4868},
}

@article{armourSeasurfaceTemperaturePattern2024,
	title = {Sea-surface temperature pattern effects have slowed global warming and biased warming-based constraints on climate sensitivity},
	volume = {121},
	issn = {0027-8424, 1091-6490},
	url = {https://pnas.org/doi/10.1073/pnas.2312093121},
	doi = {10.1073/pnas.2312093121},
	language = {en},
	number = {12},
	journal = {Proceedings of the National Academy of Sciences},
	author = {Armour, Kyle C. and Proistosescu, Cristian and Dong, Yue and Hahn, Lily C. and Blanchard-Wrigglesworth, Edward and Pauling, Andrew G. and Jnglin Wills, Robert C. and Andrews, Timothy and Stuecker, Malte F. and Po-Chedley, Stephen and Mitevski, Ivan and Forster, Piers M. and Gregory, Jonathan M.},
	month = mar,
	year = {2024},
	pages = {e2312093121},
}

@article{andrewsDependenceRadiativeForcing2015,
	title = {The {Dependence} of {Radiative} {Forcing} and {Feedback} on {Evolving} {Patterns} of {Surface} {Temperature} {Change} in {Climate} {Models}},
	volume = {28},
	issn = {0894-8755, 1520-0442},
	url = {http://journals.ametsoc.org/doi/10.1175/JCLI-D-14-00545.1},
	doi = {10.1175/JCLI-D-14-00545.1},
	language = {en},
	number = {4},
	journal = {Journal of Climate},
	author = {Andrews, Timothy and Gregory, Jonathan M. and Webb, Mark J.},
	month = feb,
	year = {2015},
	pages = {1630--1648},
}

@article{rugensteinDependenceGlobalRadiative2016,
	title = {Dependence of global radiative feedbacks on evolving patterns of surface heat fluxes},
	volume = {43},
	copyright = {http://onlinelibrary.wiley.com/termsAndConditions\#vor},
	issn = {0094-8276, 1944-8007},
	url = {https://agupubs.onlinelibrary.wiley.com/doi/10.1002/2016GL070907},
	doi = {10.1002/2016GL070907},
	language = {en},
	number = {18},
	journal = {Geophysical Research Letters},
	author = {Rugenstein, Maria A. A. and Caldeira, Ken and Knutti, Reto},
	month = sep,
	year = {2016},
	pages = {9877--9885},
}

@article{heldProbingFastSlow2010,
	title = {Probing the {Fast} and {Slow} {Components} of {Global} {Warming} by {Returning} {Abruptly} to {Preindustrial} {Forcing}},
	volume = {23},
	issn = {1520-0442, 0894-8755},
	url = {http://journals.ametsoc.org/doi/10.1175/2009JCLI3466.1},
	doi = {10.1175/2009JCLI3466.1},
	language = {en},
	number = {9},
	journal = {Journal of Climate},
	author = {Held, Isaac M. and Winton, Michael and Takahashi, Ken and Delworth, Thomas and Zeng, Fanrong and Vallis, Geoffrey K.},
	month = may,
	year = {2010},
	pages = {2418--2427},
}

@article{armourTimeVaryingClimateSensitivity2013,
	title = {Time-{Varying} {Climate} {Sensitivity} from {Regional} {Feedbacks}},
	volume = {26},
	issn = {0894-8755, 1520-0442},
	url = {http://journals.ametsoc.org/doi/10.1175/JCLI-D-12-00544.1},
	doi = {10.1175/JCLI-D-12-00544.1},
	language = {en},
	number = {13},
	journal = {Journal of Climate},
	author = {Armour, Kyle C. and Bitz, Cecilia M. and Roe, Gerard H.},
	month = jul,
	year = {2013},
	pages = {4518--4534},
}

@article{millarModifiedImpulseresponseRepresentation2017,
	title = {A modified impulse-response representation of the global near-surface air temperature and atmospheric concentration response to carbon dioxide emissions},
	volume = {17},
	issn = {1680-7324},
	url = {https://acp.copernicus.org/articles/17/7213/2017/},
	doi = {10.5194/acp-17-7213-2017},
	language = {en},
	number = {11},
	journal = {Atmospheric Chemistry and Physics},
	author = {Millar, Richard J. and Nicholls, Zebedee R. and Friedlingstein, Pierre and Allen, Myles R.},
	month = jun,
	year = {2017},
	pages = {7213--7228},
}

@article{beuschEmulatingEarthSystem2020,
	title = {Emulating {Earth} system model temperatures with {MESMER}: {From} global mean temperature trajectories to grid-point-level realizations on land},
	volume = {11},
	issn = {21904987},
	doi = {10.5194/esd-11-139-2020},
	number = {1},
	journal = {Earth System Dynamics},
	publisher = {Copernicus GmbH},
	author = {Beusch, Lea and Gudmundsson, Lukas and Seneviratne, Sonia I.},
	month = feb,
	year = {2020},
	pages = {139--159},
}

%
%
%
%
%

\end{document}